\documentclass[aps, prb, twocolumn,superscriptaddress,amsmath,amssymb,reprint]{revtex4-2}
\usepackage{graphicx}% Include figure files
\usepackage{float} %设置图片浮动位置的宏包
\usepackage{subfigure} %插入多图时用子图显示的宏包
\usepackage{dcolumn}% Align table columns on decimal point
\usepackage{bm}% bold math
\usepackage[dvipsnames]{xcolor}
\usepackage{physics}
\usepackage{pifont}
\usepackage{comment}
\usepackage{makecell}
\usepackage{svg}
\usepackage[utf8]{inputenc}
\usepackage[T1]{fontenc}

\usepackage[normalem]{ulem}
\begin{document}

\title{Spin-chirality-driven second-harmonic generation in two-dimensional magnet CrSBr}

\author{Dezhao \surname{Wu}}
\affiliation{State Key Laboratory of Low Dimensional Quantum Physics and Department of Physics, Tsinghua University, Beijing, 100084, China}

\author{Yong \surname{Xu}}
\email{yongxu@mail.tsinghua.edu.cn}
\affiliation{State Key Laboratory of Low Dimensional Quantum Physics and Department of Physics, Tsinghua University, Beijing, 100084, China}
\affiliation{Frontier Science Center for Quantum Information, Beijing, China}
\affiliation{RIKEN Center for Emergent Matter Science (CEMS), Wako, Saitama 351-0198, Japan}

\author{Meng \surname{Ye}}
\email{mye@gscaep.ac.cn}
\affiliation{Graduate School of China Academy of Engineering Physics, Beijing, 100193, China}

\author{Wenhui \surname{Duan}}
\affiliation{State Key Laboratory of Low Dimensional Quantum Physics and Department of Physics, Tsinghua University, Beijing, 100084, China}
\affiliation{Frontier Science Center for Quantum Information, Beijing, China}
\affiliation{Institute for Advanced Study, Tsinghua University, Beijing 100084, China}

% \date{\today}
\begin{abstract}
The interplay between magnetism and light can create abundant optical phenomena. Here, we demonstrate the emergence of an unconventional magnetization-induced second-harmonic generation (MSHG) stemming from vector spin chirality, denoted as chiral second-harmonic generation (SHG). Taking the antiferromagnetic (AFM) CrSBr bilayer as a prototype, we theoretically show that, via spin canting, the chiral SHG can be continuously tuned from zero to a value one order of magnitude larger than its intrinsic MSHG. Remarkably, chiral SHG is found to be proportional to spin chirality and spin-canting-induced electric polarization, while intrinsic MSHG is proportional to the N\'{e}el vector, demonstrating their different physical mechanisms. Additionally, we reveal a unique interference effect between these two types of MSHG under the reversal of spin-canting direction, generating a giant modulation of SHG signals. Our work not only uncovers a unique SHG with exceptional tunability but also promotes the applications of AFM optical devices and magnetoelectric detection techniques.
\end{abstract}

\keywords{Suggested keywords}

\maketitle

%=================================
\section*{Introduction} 
%=================================
Light-matter interactions in magnetic materials can create rich physics and have important applications. 
Second-harmonic generation (SHG) is a frequency-doubling effect that occurs when light passes through a material without inversion ($\mathcal{P}$) symmetry. 
Recently, giant magnetization-induced SHG (MSHG) has been discovered in van der Waals magnets with a centrosymmetric crystal structure, in which the $\mathcal{P}$ symmetry is broken solely by antiferromagnetic (AFM) ordering \cite{Wu_CrI3_2019,MnPS3_2020_prl,zhu_2021_nanoletters_CrSBr}. 
A prominent feature of MSHG observed in those experiments is its time-reversal ($\mathcal{T}$) odd (also called $c$-type) nature \cite{Wu_CrI3_2019, MnPS3_2020_prl, zhu_2021_nanoletters_CrSBr, npj, 2024-pnas-NLMO, Toyoda-2023-prm, Wu-2023-ACSnano, Tokura_FeGaO3_2004}, in contrast to the commonly observed $\mathcal{T}$-even (also called $i$-type) crystal SHG stemming from the noncentrosymmetric crystal structure. 
From an application perspective, MSHG offers multiple capabilities.
Firstly, it can act as a sensitive probe for discerning subtle magnetic configurations \cite{Wu_CrI3_2019,MnPS3_2020_prl,zhu_2021_nanoletters_CrSBr}, detecting magnetic transitions \cite{Wu_CrI3_2019, MnPS3_2020_prl, zhu_2021_nanoletters_CrSBr}, and analyzing magnetoelectric (ME) coupling \cite{MnPS3_2020_prl}. 
Secondly, during magnetization reversal,
MSHG can interfere with crystal SHG  \cite{npj, 2024-pnas-NLMO, Toyoda-2023-prm,Wu-2023-ACSnano,Tokura_FeGaO3_2004}, thereby enabling a wide array of modulations of SHG signals. 
However, notable MSHG is relatively rare and has been experimentally observed in only a restricted set of two-dimensional (2D) AFM materials \cite{Wu_CrI3_2019, MnPS3_2020_prl, zhu_2021_nanoletters_CrSBr}. Therefore, the realization of large and tunable MSHG through alternative mechanisms is of great interest in both fundamental research domains and device-application scenarios.

In contrast to collinear AFM orderings, vector spin chirality $\boldsymbol{\kappa} = \mathbf{S_A} \times \mathbf{S_B}$ arising from the noncollinear arrangement of spins $\mathbf{S_A}$ and $\mathbf{S_B}$ provides new possibilities to achieve large and tunable MSHG. 
In one aspect, $\boldsymbol{\kappa}$ has been proven to be responsible for abundant responses that emerge in noncollinear spin systems, such as chiral Hall effects \cite{chiral_Hall_2020_prl,chiral_Hall_2021_communications_physics,Yao_2020_prm_chiral-AHE&ANE,chiral_Hall_2022_npjQM_MBT,Yao_2023_In-plane-anomalous-Hall,Yao_2023_nanolett_chiral-QAH&QTH}, chiral magneto-optical effects \cite{chiral_Hall_2021_communications_physics}, and chiral photocurrents \cite{chiral_photocurrents_2023_APL}. 
The common feature of these chiral responses is their sensitivity to spin rotation \cite{chiral_Hall_2020_prl, chiral_Hall_2021_communications_physics, Yao_2020_prm_chiral-AHE&ANE, chiral_Hall_2022_npjQM_MBT, Yao_2023_In-plane-anomalous-Hall, Yao_2023_nanolett_chiral-QAH&QTH, chiral_photocurrents_2023_APL}, suggesting that these responses can be fine-tuned by changes in the sign and magnitude of $\boldsymbol{\kappa}$.
Besides, they can be easily achieved in AFM systems through magnetic-field-induced spin canting \cite{chiral_Hall_2021_communications_physics,Yao_2020_prm_chiral-AHE&ANE,chiral_Hall_2022_npjQM_MBT,Yao_2023_In-plane-anomalous-Hall,chiral_photocurrents_2023_APL}. 
In another aspect, notable MSHG has been observed experimentally in spin-spiral NiI$_2$ \cite{NiI2_2021_nl,NiI2_2022_Nature} and spin-canted CuCrP$_2$S$_6$ \cite{2024_AM_chiral-SHG-in-CuCrP2S6}.
These experimental evidences have enabled the unique characterization of type-II multiferroics \cite{NiI2_2021_nl,NiI2_2022_Nature} and the giant modulation of SHG signals \cite{2024_AM_chiral-SHG-in-CuCrP2S6}.
Moreover, it offers valuable insights into the unconventional origins and properties of MSHG in noncollinear magnets with non-vanishing $\boldsymbol{\kappa}$. 
Therefore, it is of great interest to investigate the existence of {\it chiral SHG} driven by $\boldsymbol{\kappa}$ and to explore its unique properties, distinct from traditional $c$-type MSHG.

The newly discovered 2D AFM semiconductor CrSBr \cite{2020_AM_CrSBr,zhu_2021_nanoletters_CrSBr,zhu_2021_nature-mater_CrSBr,xu_2022_nature-nanotech_CrSBr,zhu_2022_nature_CrSBr,wang_2022_acsnano_CrSBr,xu_2023_nature-nanotech_CrSBr_tunable_exciton-magnon_coupling,2023_Nature_CrSBr_tunable_MO,2024-NC-CrSBr-doping,zhu_2024_CrSBr_nanoletter_review} provides an ideal platform to explore the concept of chiral SHG. 
First, its centrosymmetric crystal structure ensures the absence of signals from the crystal SHG \cite{zhu_2021_nanoletters_CrSBr}. 
Secondly, the magnetic configuration of CrSBr can be easily tuned to a canted AFM (cAFM) state at an arbitrary spin-canting angle with a small magnetic field ($<$ 2\,T) \cite{zhu_2021_nature-mater_CrSBr,2024-NC-CrSBr-doping,zhu_2024_CrSBr_nanoletter_review,2023_Nature_CrSBr_tunable_MO}, imprinting a flexible $\boldsymbol{\kappa}$. The observations of the photoluminescence \cite{zhu_2021_nature-mater_CrSBr,xu_2022_nature-nanotech_CrSBr,2023_Nature_CrSBr_tunable_MO} and reflective magnetic circular dichroism signals \cite{zhu_2021_nature-mater_CrSBr,xu_2022_nature-nanotech_CrSBr} in experiments verified that the spin canting states can be reliably induced and maintained. More importantly, CrSBr exhibits strong coupling between the canting angle and the exciton \cite{zhu_2021_nature-mater_CrSBr,xu_2022_nature-nanotech_CrSBr}, magnon \cite{zhu_2022_nature_CrSBr,xu_2023_nature-nanotech_CrSBr_tunable_exciton-magnon_coupling}, and polariton \cite{2023_Nature_CrSBr_tunable_MO} states, giving rise to a diverse range of highly tunable properties. These exciting experimental observations suggest the strong coupling between $\boldsymbol{\kappa}$ and MSHG in cAFM CrSBr. 

In this work, through symmetry analysis and first-principles calculations \cite{Wang_2017,Chen_2022}, we uncovered the existence of chiral SHG in CrSBr, which is $i$-type and fundamentally different from the commonly observed $c$-type MSHG. We further found that, via spin canting, the chiral SHG in CrSBr can be continuously tuned from zero to a large value, one order of magnitude larger than its intrinsic MSHG. The band-nesting effect and anisotropic optical absorption of CrSBr are found to account for the large and anisotropic value of chiral SHG.  
Remarkably, we discovered that the chiral SHG is proportional to $\boldsymbol{\kappa}$ and the spin-canting-induced electric polarization, while the intrinsic MSHG is proportional to the N\'{e}el vector.
These observations enable the detection of $\boldsymbol{\kappa}$ and the N\'{e}el vector, as well as the characterization of ME effects by measuring chiral SHG and intrinsic MSHG.
Furthermore, we revealed that the chiral SHG can interfere with the intrinsic MSHG under the reversal of spin-canting direction, which dramatically modifies the polarization and strength of the SHG signals. 
Our work not only uncovers a mechanism to generate MSHG with superb tunability but also facilitates fundamental research on ME coupling and device applications in widespread AFM materials.

%=================================
\section*{Results} \label{Results}
%=================================
%--------------------------------------------------
 \subsection*{General theory of chiral SHG}
%--------------------------------------------------
%******************************************************************%
\begin{figure*}
    \includegraphics[width=0.7\linewidth]{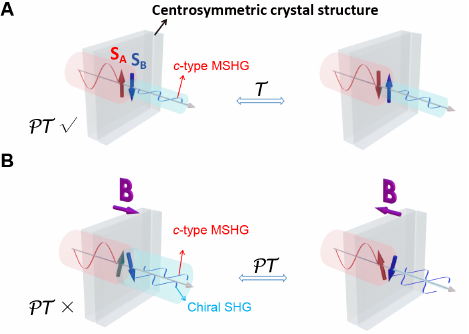}
    \caption{\textbf{Concepts of chiral SHG and its interference effects.} 
    (A) Schematics of SHG generated in a BL AFM material with $\mathcal{PT}$ symmetry (left panel) and the influence of $\mathcal{T}$ operation (right panel). (B) Schematics of chiral SHG and vector spin chirality $\boldsymbol{\kappa} = \mathbf{S_A} \times \mathbf{S_B}$ generated in the same material under a magnetic field $\mathbf{B}$ perpendicular to the direction of the N\'{e}el vector and the influence of $\mathcal{PT}$ operation. The $c$-type MSHG is unaffected by $\mathbf{B}$ field reversal while the chiral SHG changes the sign. Thus the two terms can constructively (left panel) or destructively (right panel) interfere with each other for the positive and negative $\mathbf{B}$ field directions, respectively. The red and blue arrows represent magnetic moments in each layer.} 
 \label{Fig1}
\end{figure*}
%******************************************************************%
Chiral SHG generally belongs to a unique class of $i$-type MSHG, which is demonstrated in the classification of electric-dipole SHG (ED-SHG) in Table \ref{tab:classification}. When the $\mathcal{P}$ symmetry is broken by a noncentrosymmetric crystal structure, the accompanying breaking of space-time-reversal ($\mathcal{PT}$) symmetry can give rise to the crystal SHG \cite{NbOCl2_2023_nature,Yao_2021_SA_twisted-hBN,wang_2024_prm} which is $i$-type. 
On the other hand, if the $\mathcal{P}$ symmetry is solely broken by a noncentrosymmetric magnetic structure, the accompanying breaking of the $\mathcal{T}$ symmetry can give rise to $c$-type MSHG \cite{Wu_CrI3_2019,MnPS3_2020_prl,zhu_2021_nanoletters_CrSBr,npj,2024-pnas-NLMO,Toyoda-2023-prm,Wu-2023-ACSnano,Tokura_FeGaO3_2004, Birss_1964, shen1984principles}.
However, complex magnetic structures, such as noncollinear or noncoplanar magnetic orderings, can simultaneously break both $\mathcal{P}$ and $\mathcal{PT}$ symmetries, and therefore they can induce additional $i$-type SHG that originates from magnetic ordering (details in section S3.1 \cite{SI}). As summarized in Table \ref{tab:classification}, the chiral SHG generally belongs to this $i$-type MSHG, setting it apart from the well-studied $i$-type crystal SHG \cite{NbOCl2_2023_nature,Yao_2021_SA_twisted-hBN,wang_2024_prm} and $c$-type MSHG \cite{Wu_CrI3_2019,MnPS3_2020_prl,zhu_2021_nanoletters_CrSBr,npj,2024-pnas-NLMO,Toyoda-2023-prm,Wu-2023-ACSnano,Tokura_FeGaO3_2004,Birss_1964,shen1984principles}. In fact, this $i$-type MSHG has long been overlooked in previous studies, and its properties are rarely investigated \cite{Xiao_classification}.

Chiral SHG is the most easily achieved case among this unique class of $i$-type MSHG, as it can be induced by spin canting in bipartite AFM materials by applying a magnetic field. 
In Fig. \ref{Fig1}A, a bilayer (BL) A-type AFM material where the $\mathcal{P}$ symmetry is purely broken by the AFM ordering (having $\mathcal{PT}$ symmetry) exhibits pure $c$-type MSHG with magnetic origin.
As illustrated in Fig. \ref{Fig1}B, when a magnetic field $\mathbf{B}$ perpendicular to the N\'{e}el vector $\mathbf{L}$ is applied, the magnetic moments in each layer ($\mathbf{S_A}$ and $\mathbf{S_B}$) cant towards the external field, and a nonzero vector spin chirality $\boldsymbol{\kappa} = \mathbf{S_A} \times \mathbf{S_B}$ emerges. Generally, $\boldsymbol{\kappa}$ is present as long as $\mathbf{B}$ is not parallel to $\mathbf{L}$. As $\boldsymbol{\kappa}$ breaks both the $\mathcal{P}$ and $\mathcal{PT}$ symmetries (depicted in fig. S19 \cite{SI}), it induces an additional chiral SHG.  

%Advantages of the chiral SHG
The flexibility and magnetic origin inherent in $\boldsymbol{\kappa}$ suggest that the chiral SHG would behave differently from either $c$-type MSHG or $i$-type crystal SHG.
On the one hand, while chiral SHG and $c$-type MSHG share the same magnetic origin, their behaviors diverge under magnetic tuning and $\mathcal{T}$ operation.
On the other hand, while chiral SHG shares the same $i$-type nature as crystal SHG, it is anticipated to be much more tunable due to the greater flexibility of $\boldsymbol{\kappa}$ compared to the relatively rigid crystal structure.

%~~~~~~~~~~~~~~~~~~~~~~~~~~~~~~~~~~~~~~~~~~~~~~~~~~~~~~~~~~~~~~~~~~%
\begin{table} 
    \centering
    \small % Set the font size for the whole table
    \renewcommand\theadfont{\small} % Ensure \thead font size matches
    \begin{tabular}{ccc}
        \hline \hline 
         \thead{ED-SHG\\($\mathcal{P}$ broken)}& \thead{Crystal structure\\origin} & \thead{Magnetic structure\\origin}\\ \hline 
         \thead{$i$-type\\($\mathcal{PT}$ broken)} & Crystal SHG & Chiral SHG \\ \hline 
         \thead{$c$-type\\($\mathcal{T}$ broken)}  &  NA & $c$-type MSHG \\ \hline \hline 
    \end{tabular}
    \caption{\textbf{Classification of ED-SHG.} $\mathcal{P}$, $\mathcal{T}$, and $\mathcal{PT}$ are inversion symmetry, time-reversal symmetry, and space-time-reversal symmetry. NA, not applicable.}
    \label{tab:classification}
\end{table}
%~~~~~~~~~~~~~~~~~~~~~~~~~~~~~~~~~~~~~~~~~~~~~~~~~~~~~~~~~~~~~~~~~~
%NLMO
More importantly, chiral SHG can induce a unique interference effect. Conventionally, interference occurs between crystal SHG and $c$-type MSHG when the $\mathbf{B}$ field {\it parallel} to the magnetic easy axis is reversed. Consequently, it reverses the magnetization and acts like a $\mathcal{T}$ operation, keeping the sign of the $i$-type crystal SHG susceptibility $\chi^{abc}_{\rm crystal}$ unchanged while reversing the sign of the $c$-type MSHG susceptibility $\chi^{abc}_c$ as \cite{npj, 2024-pnas-NLMO, Toyoda-2023-prm,Wu-2023-ACSnano,Tokura_FeGaO3_2004} 
\begin{equation}
    P^a(2\omega)\propto(\chi^{abc}_{\rm crystal}\pm\chi^{abc}_c)E^b(\omega)E^c(\omega)\,.
\end{equation}
Here, $E(\omega)$ is the electric field of incident light, and $P(2\omega)$ is the nonlinear polarization, with $a$, $b$, and $c$ representing the Cartesian coordinates. 
The Einstein summation convention is adopted in all formulas for repeated indices, and the shorthand notation $\chi^{abc}$ is adopted to represent $\chi^{abc}(2\omega;\omega,\omega)$ in the rest of the article.  
The $\pm$ represents the influence of the $\mathcal{T}$ operation, and the interference is reflected by the change in the intensity of the SHG $I(2\omega)\propto |P(2\omega)|^2$.
In contrast, in $\mathcal{PT}$ symmetric magnets with a centrosymmetric crystal structure depicted in Fig. \ref{Fig1}A, the crystal SHG vanishes, and only $c$-type MSHG survives. 
However, a unique interference can be triggered by introducing a $\mathbf{B}$ field {\it not parallel to} the magnetic easy axis, as shown in Fig. \ref{Fig1}B. 
Notably, flipping $\mathbf{B}$ reverses spin-canting direction and $\boldsymbol{\kappa}$, acting as a $\mathcal{PT}$ operation, and results in the sign reversal of the $i$-type chiral SHG susceptibility $\chi^{abc}_{\rm chiral}$ as
\begin{equation}
    P^a(2\omega)\propto(\pm\chi^{abc}_{\rm chiral}+\chi^{abc}_c)E^b(\omega)E^c(\omega)\,,
\end{equation}\label{eq:chiral-NLMO}
where the $\pm$ represents the influence of $\mathcal{PT}$ operation. 
In addition to differences in magnetic field direction and the symmetry operations involved, in this unique interference effect, the tunability of the vector $\boldsymbol{\kappa}$ enables the continuous modulation of $\chi_{\rm chiral}$, leading to a precisely tunable interference strength under the reversal of $\boldsymbol{\kappa}$ at an arbitrary $\omega$. This is much more flexible than conventional interference effects that depend on the rigid 180$^\circ$ reversal of the spin.
It is worth noting that this tunability of interference is also applicable between chiral and crystal SHG. 
Furthermore, in AFM magnets with fully compensated magnetization, the $\mathbf{B}$ field can easily manipulate the vector $\boldsymbol{\kappa}$ and $\boldsymbol{\kappa}$ reversals, while flipping  the N\'{e}el vector $\mathbf{L}$ is challenging.
The above features enable the unique interference effects under $\boldsymbol{\kappa}$ reversal to be realized in a variety of AFM materials.

%--------------------------------------------------
 \subsection*{Large and anisotropic chiral SHG in CrSBr}
%--------------------------------------------------
%******************************************************************%
\begin{figure*}
    \includegraphics[width=\linewidth]{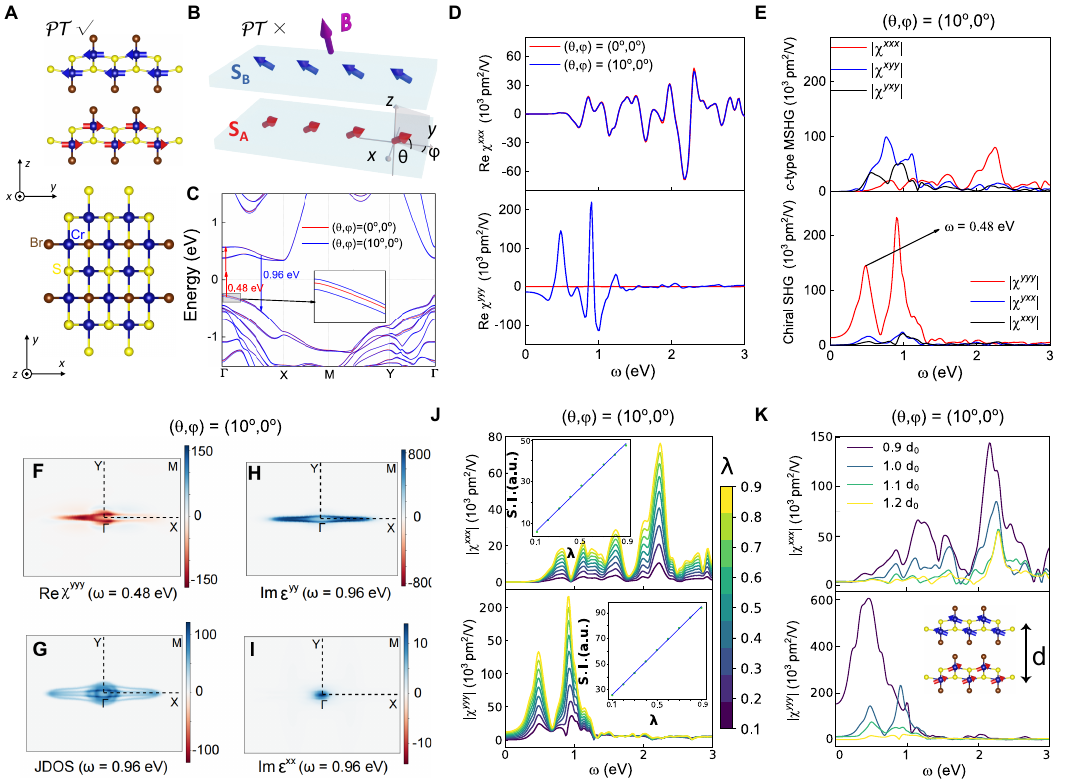} 
    \caption{\textbf{Large and anisotropic chiral SHG in BL CrSBr.} (A) Crystallographic and magnetic structures of BL CrSBr. The side view is in the upper panel while the top view is in the lower panel. (B) Schematics of $\mathcal{PT}$-symmetry broken by a magnetic field $\mathbf{B}$ perpendicular to N\'{e}el vector and the definition of canting angle $(\theta,\varphi)$ at spin sublattice A. (C) spin-canting-induced changes on the band structure of BL CrSBr. (D) spin-canting-induced changes on real parts of SHG susceptibilities $\chi^{xxx}$ and $\chi^{yyy}$ of BL CrSBr. (E) Norm of all in-plane SHG susceptibilities at (10$^\circ$,0$^\circ$) canting state. The upper panel is for the $c$-type MSHG components and the lower panel is for the chiral SHG components. (F) The real part of $\mathbf{k}$-resolved chiral component $\chi^{yyy}$ with the incident photon energy at 0.48\,eV. (G) $\mathbf{k}$-resolved JDOS, (H) optical absorption Im\,$\epsilon^{yy}$, and (I) Im\,$\epsilon^{xx}$ with the incident photon energy at 0.96\,eV at (10$^\circ$,0$^\circ$) canting state. (J) The influence of SOC magnitude $\lambda$ and (K) and interlayer distance $d$ on $|\chi^{xxx}|$ and $|\chi^{yyy}|$ at (10$^\circ$,0$^\circ$) canting state. The insets of (J) represent spectral integrals of corresponding SHG susceptibilities. a.u., arbitrary units. The $d_0$ in (K) represents the interlayer distance of BL CrSBr at equilibrium. } \label{Fig2}
\end{figure*}
%******************************************************************%
%~~~~~~~~~~~~~~~~~~~~~~~~~~~~~~~~~~~~~~~~~~~~~~~~~~~~~~~~~~~~~~~~~~%
\begin{table*} 
    \centering
    \small % 设置整个表格的字体大小为 small
    \begin{tabular}{ccccc}
        \hline \hline 
         & ($0^\circ$,$0^\circ$) & ($\theta\neq 0^\circ$,$0^\circ$) & ($0^\circ$,$\varphi\neq 0^\circ$) & ($\theta\neq 0^\circ$,$\varphi\neq 0^\circ$) \\ \hline
         MPGs  & $m'mm$ & $m'2'm$ &   $22'2'$ & $2'$ \\ \hline 
         Chiral SHG  & NA & $yyy,yxx,xxy$ & NA & $yyy,yxx,xxy$ \\ \hline 
         $c$-type MSHG &  $xxx,xyy,yxy$ & $xxx,xyy,yxy$ & $xxx,xyy,yxy$ & $xxx,xyy,yxy$ \\ \hline \hline
    \end{tabular}
    \caption{\textbf{Symmetry of BL CrSBr.} Magnetic point groups (MPGs) of BL CrSBr with different spin-canting configurations and their corresponding nonzero SHG components. The notations of MPGs follow the direction sequence of $\hat{x}$, $\hat{y}$ and $\hat{z}$. NA, not applicable.} 
    % \MY{The first one is 2x/Mx' 2y'/My 2z'/Mz. The second one is Mx' 2y' Mz; The third one is 2x 2y' 2z'; The fourth one is 2y'} 
	\label{tab:sym}
\end{table*}
%~~~~~~~~~~~~~~~~~~~~~~~~~~~~~~~~~~~~~~~~~~~~~~~~~~~~~~~~~~~~~~~~~~%
%structure and canting effects of CrSBr
The newly discovered 2D magnetic semiconductor CrSBr is an ideal candidate to explore chiral SHG. 
In addition to its remarkable air stability and high N\'{e}el temperature (150\,K), its magnetic structure can easily be adjusted to a cAFM state at an arbitrary canting angle under a magnetic field of up to 2\,T \cite{zhu_2021_nature-mater_CrSBr,zhu_2021_nanoletters_CrSBr,2024-NC-CrSBr-doping,zhu_2024_CrSBr_nanoletter_review,xu_2022_nature-nanotech_CrSBr,zhu_2022_nature_CrSBr,xu_2023_nature-nanotech_CrSBr_tunable_exciton-magnon_coupling,2023_Nature_CrSBr_tunable_MO,2020_AM_CrSBr,wang_2022_acsnano_CrSBr}, giving rise to highly tunable $\boldsymbol{\kappa}$.
As depicted in Fig. \ref{Fig2}A, the AFM phase of BL CrSBr belongs to the magnetic point group $m^\prime mm$, with the N\'{e}el vector $\mathbf{L}$ aligns along the $\hat{y}$-axis. 
In this phase, the $\mathcal{P}$ and $\mathcal{T}$ symmetries are broken, while $\mathcal{PT}$ symmetry is preserved. 
Fig. \ref{Fig2}B illustrates the spin canting and $\mathcal{PT}$-symmetry broken when a $\mathbf{B}$ field perpendicular to $\mathbf{L}$ is applied, inducing a nonzero $\boldsymbol{\kappa}$.
The canting states of the BL CrSBr are represented by the canting angles of $\mathbf{S_A}$, i.e., ($\theta, \varphi$) defined in Fig. \ref{Fig2}B, where $\theta \in [-90^\circ, 90^\circ]$ is the polar angle between the $xy$-plane, and $\varphi \in [0^\circ, 360^\circ)$ is the azimuth angle with respect to the positive $\hat{y}$-axis. 
The canting angles of $\mathbf{S_B}$ are ($\theta,180^\circ -\varphi$). Accordingly, the N\'{e}el vector $\mathbf{L}$, the net magnetization $\mathbf{M}$, and the vector spin chirality $\boldsymbol{\kappa}$ are expressed as
\begin{equation}
    \begin{aligned}
    \mathbf{L} &= \mathbf{S_A} - \mathbf{S_B} \propto (0, \cos\theta\cos\varphi, 0) \,, \\
    \mathbf{M} &= \mathbf{S_A} + \mathbf{S_B} \propto (\cos\theta\sin\varphi, 0, \sin\theta) \,,\\
    \boldsymbol{\kappa}&= \mathbf{S_A} \times \mathbf{S_B} \propto (\sin2\theta\cos\varphi, 0, -\cos^2\theta \sin2\varphi).
    \end{aligned}\label{order-parameter}
\end{equation}
The magnetic point groups and corresponding $i$/$c$-type tensor components at different canting configurations are summarized in Table \ref{tab:sym}. 
Our setting of $\mathbf{B} \perp \mathbf{L}$ is consistent with experiments
\cite{2020_AM_CrSBr,zhu_2021_nature-mater_CrSBr,xu_2022_nature-nanotech_CrSBr,zhu_2022_nature_CrSBr,wang_2022_acsnano_CrSBr}, and the $\mathcal{C}_{2y}\mathcal{T}$ symmetry that persists in this case always separates the $i$/$c$-type tensor components, as shown in Table \ref{tab:sym}. 
For a magnetic field $\mathbf{B}$ with an arbitrary direction, an additional set of $(\theta',\varphi')$ for $\mathbf{S_B}$ needs to be introduced, and the $i$/$c$-type MSHG can be mixed with each other in a single tensor component. 
%\textcolor{black}{(Even in this case, if SA and SB are coplaner in yz-plane, the MyT can also separete the i/c-type tensors.)}

As the $\mathbf{B}$ field is usually applied along the $\hat{z}$-axis of CrSBr in experiment \cite{zhu_2022_nature_CrSBr, xu_2022_nature-nanotech_CrSBr, zhu_2021_nature-mater_CrSBr, 2020_AM_CrSBr, wang_2022_acsnano_CrSBr} , we first focused on the corresponding canting states ($\theta \neq 0^\circ ,\varphi = 0^\circ$).
In this case, $\mathcal{PT}$, $\mathcal{C}_{2z}\mathcal{T}$, $\mathcal{C}_{2x}$, and $\mathcal{M}_{y}$ symmetries are broken, resulting in the emergence of $i$-type tensor components $\chi^{yyy}$, $\chi^{yxx}$, and $\chi^{xxy}$, as shown in Table \ref{tab:sym}.  
For a small canting angle (10$^\circ$,0$^\circ$) which corresponds to a weak $\mathbf{B}$ field of approximately 0.3\,T along the $\hat{z}$-axis \cite{zhu_2021_nature-mater_CrSBr,zhu_2024_CrSBr_nanoletter_review}, while the band degeneracy previously protected by $\mathcal{PT}$ symmetry is only slightly lifted (see Fig. \ref{Fig2}C), the SHG susceptibilities change notably. 
As shown in Fig. \ref{Fig2}D, while the $c$-type $\chi^{xxx}$ component remains nearly unchanged, a large $i$-type $\chi^{yyy}$ signal emerges, with a considerable value in the static limit, demonstrating the chiral SHG induced by $\boldsymbol{\kappa}$. 
Within these newly emerging chiral SHG components, $\chi^{yyy}$ is larger in magnitude than the intrinsic $c$-type MSHG, even at the small canting angle $\theta = 10^\circ$, while the other chiral SHG components have a magnitude similar to the intrinsic $c$-type MSHG (see Fig. \ref{Fig2}E).
This unusual anisotropy of chiral SHG is strongly correlated with the calculated magnetization-induced ferroelectricity \cite{KNB,gKNB} which is along the $\hat{y}$-direction, with electronic polarization $\pm$0.0225 pC/m at ($\pm$10$^\circ$,0$^\circ$) canting states. 
Moreover, this finding is also in agreement with previously reported enhanced SHG in the polarization direction \cite{Huang_2023_NC_NbOCl,2024_AM_chiral-SHG-in-CuCrP2S6}.

To gain a more in-depth understanding of the large and anisotropic chiral SHG in CrSBr, we performed an analysis of the microscopic origin of the 0.48\,eV peak in $|\chi^{yyy}|$ (see Fig. \ref{Fig2}E). 
As shown in Fig. \ref{Fig2}F, $\mathbf{k}$-resolved chiral SHG susceptibility $\chi^{yyy}$ at 0.48\,eV exhibits a concentrated contribution near $\Gamma$ and along the $\Gamma$-X direction.
Since 0.48\,eV is even smaller than the band-gap energy, this peak is dominated by the double-photon-resonance process. As SHG is generally related to the joint density of states (JDOS) and the linear optical absorption strength $\mathrm{Im} \epsilon^{aa}$ (see expressions in section S3.3 \cite{SI}) \cite{Huang_2023_NC_NbOCl, 2013_prb_band-nesting}, we calculated the $\mathbf{k}$-resolved JDOS and $\mathrm{Im} \epsilon^{aa}$ at 0.96\,eV. As shown in Fig. \ref{Fig2}G, a large anisotropic JDOS concentrating along $\Gamma$-X was found, which originates from the nearly parallel conduction and valence bands along the $\Gamma$-X direction around the band gap, known as the band-nesting effect \cite{Huang_2023_NC_NbOCl,2020_nanolett_band-nesting,2013_prb_band-nesting}, as depicted in Fig. \ref{Fig2}C.
The anisotropic distribution of $\mathbf{k}$-resolved JDOS and the distinct band dispersion between $\Gamma$-X and $\Gamma$-Y directions reveal the quasi-1D character of CrSBr \cite{CrSBr_2023_1D_ACS-nano,zhu_2024_CrSBr_nanoletter_review}, and this quasi-1D feature is also reflected in the anisotropy of optical absorption strength. 
As shown in Figs. \ref{Fig2} (H and I), the Im\,$\epsilon^{yy}$ is much larger than Im\,$\epsilon^{xx}$ at 0.96\,eV and shows a similar concentrated contribution along $\Gamma$-X as $\chi^{yyy}$ and JDOS.
Consequently, the band-nesting effect and the strong optical absorption strength Im\,$\epsilon^{yy}$ act cooperatively and result in the large $|\chi^{yyy}|$.

%******************************************************************%
\begin{figure*}
    \includegraphics[width=\linewidth]{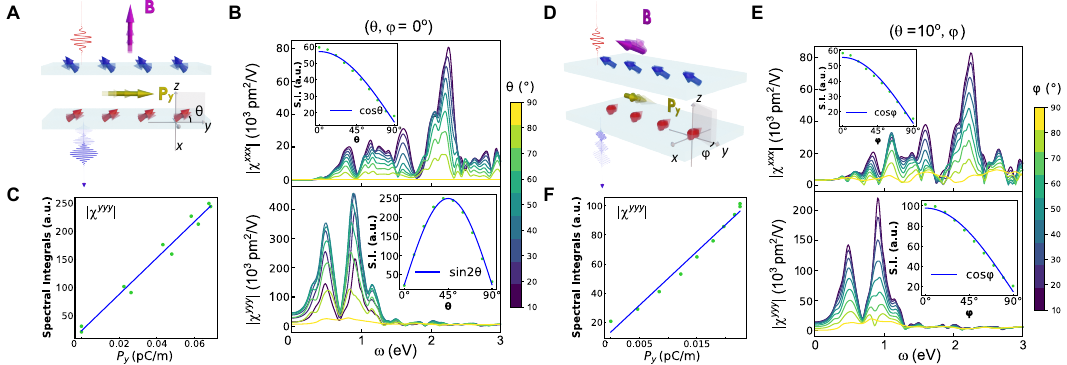}
    \caption{\textbf{Tunability of chiral SHG and ME detection in BL cAFM CrSBr.} (A) Schematics of the real-time tuning of chiral SHG under the $\hat{y}$-polarized incident light and the detection of spin-canting-induced electric polarization dynamics through the corresponding chiral SHG intensity $\propto|\chi^{yyy}|^2$. (B) The $c$-type MSHG susceptibility $|\chi^{xxx}|$ and chiral SHG susceptibility $|\chi^{yyy}|$ and (C) the relationship between the spectral integrals of $|\chi^{yyy}|$ with spin-canting-induced electric polarization under the variation of $\theta$ within the canting configuration ($\theta$, $\varphi=0^\circ$), with $\mathbf{B}$ along the $\hat{z}$-axis. (D to F) The counterparts of (A to C) under the variation of $\varphi$ within the canting configuration ($\theta=10^\circ$,$\varphi$), with $\mathbf{B}$ in the $xz$-plane. The insets of (B) and (E) represent spectral integrals of corresponding SHG susceptibilities}   \label{Fig3}
\end{figure*}
%******************************************************************%
To gain a deeper understanding of the underlying driving mechanism of the chiral SHG in CrSBr, we investigated the influence of spin-orbit coupling (SOC) and the distance between layers on the chiral SHG susceptibility $\chi^{yyy}$, using the $c$-type MSHG susceptibility $\chi^{xxx}$ as a reference. 
In addition to assessing the frequency-dependent SHG susceptibility, we also evaluated their spectral integrals $\int |\chi^{abc}(\omega)| d\omega$ (S.I.), which serves as a more intuitive and coarse-grained indicator.
As shown in \textcolor{black}{Fig. \ref{Fig2}J}, the SHG spectrum and S.I. strongly indicate that spin-canting-induced $i$-type chiral SHG susceptibility $\chi^{yyy}$, as well as intrinsic $c$-type MSHG susceptibility $\chi^{xxx}$, are strictly proportional to the magnitude of SOC, i.e. $\lambda$. 
Moreover, a negative dependence on interlayer spacing $d$ for both types of SHG are observed in \textcolor{black}{Fig. \ref{Fig2}K}, suggesting the important role of interlayer coupling in generating these SHG signals (see the corresponding S.I. in fig. S4 \cite{SI}). 
These behaviors clearly reveal that the chiral SHG in BL CrSBr originates from SOC and interlayer coupling, which is very similar to the intrinsic $c$-type MSHG in BL CrSBr. 
In contrast, the crystal SHG is less sensitive to SOC, as reported in Ref. \cite{npj}. 
Therefore, despite the fact that chiral SHG falls into the $i$-type category similar to crystal SHG, this disparity underscores their magnetic and crystallographic origins, respectively.
Moreover, both $c$-type and chiral SHG are found to be insensitive to the layer number in even-layer CrSBr. Even so, we proposed BL CrSBr as the optimal structure for observing either chiral SHG or $c$-type MSHG. 
This is because, as the number of layers increases, the phase-matching problem worsens \cite{Boyd_2020_nonlinear-opticals}, and the influence of magnetic polymorph effects is amplified \cite{Wu_2025_Nat-Mater_CrSBr-polymorphs}. (see analysis in section S1.1 \cite{SI}). In addition, all ED-SHG responses are absent in odd layers due to the presence of $\mathcal{P}$ symmetry.

%--------------------------------------------------
 \subsection*{Highly tunable chiral SHG and ME detection in CrSBr}
%--------------------------------------------------
We continuously varied the canting angles ($\theta, \varphi$) to explore the tunability of chiral SHG, simulating the effect of changes in the magnetic field $\mathbf{B}$ in experiments
\cite{2020_AM_CrSBr,zhu_2021_nanoletters_CrSBr,zhu_2021_nature-mater_CrSBr,xu_2022_nature-nanotech_CrSBr,zhu_2022_nature_CrSBr,wang_2022_acsnano_CrSBr}.
Firstly, we fixed the azimuth angle to $\varphi$=0$^\circ$ and increased the polar angle $\theta$ from 0$^\circ$ to 90$^\circ$ to simulate the influence of an increasing $\mathbf{B}$ field along the $\hat{z}$-axis, as depicted in Fig. \ref{Fig3}A. 
Figure \ref{Fig3}B shows that the chiral $|\chi^{yyy}|$ exhibits an interesting trend of initial increase followed by decrease and reaches maximum at $\theta$ = 45$^\circ$. 
The zero value at $\theta$=0$^\circ$/90$^\circ$ are forced by $\mathcal{PT}$/$\mathcal{P}$ symmetry. On the contrary, the $c$-type $|\chi^{xxx}|$ monotonously decreases to zero. It is worth noting that the maximum of the chiral $|\chi^{yyy}|$ can be one order of magnitude larger than that of the $c$-type $|\chi^{xxx}|$. More interestingly, A further S.I. analysis shows that the chiral $|\chi^{yyy}| \propto \sin{2\theta}$, while the $c$-type $|\chi^{xxx}| \propto \cos{\theta}$ (see insets of Fig. \ref{Fig3}B). 
Secondly, we fixed $\theta$ = 10$^\circ$ and magnified $\varphi$ from 0$^\circ$ to 90$^\circ$ to simulate a $\mathbf{B}$ field with the out-of-plane magnitude $B_z$ fixed and the in-plane magnitude $B_x$ increasing, as depicted in Fig. \ref{Fig3}D. Here, a finite $\theta$ is necessary to activate the nonzero chiral components $\chi^{yyy}, \chi^{yxx}$ and $\chi^{xxy}$ by breaking the $\mathcal{C}_{2z}\mathcal{T}$ and $\mathcal{C}_{2x}$ symmetries, as shown in Table \ref{tab:sym}. Interestingly, both chiral $|\chi^{yyy}|$ and $c$-type $|\chi^{xxx}|$ monotonically decrease to zero and are proportional to $\cos{\varphi}$, as illustrated in Fig. \ref{Fig3}E. 
The vanishing SHG at $\varphi=90^\circ$ is due to the restored $\mathcal{P}$ symmetry. 
In addition, the canting angle dependencies mentioned above are also observed in other chiral and $c$-type tensor components (see figs. S8 and S9 \cite{SI}). 
These features enable the precise control of chiral SHG through an external magnetic field, as depicted in Figs. \ref{Fig3} (A and D), which is superior to previously proposed strategies involving twisting \cite{Yao_2021_SA_twisted-hBN} and stacking engineering \cite{NbOCl2_2023_nature, npj}.

\begin{figure*}
    \includegraphics[width=1.0\linewidth]{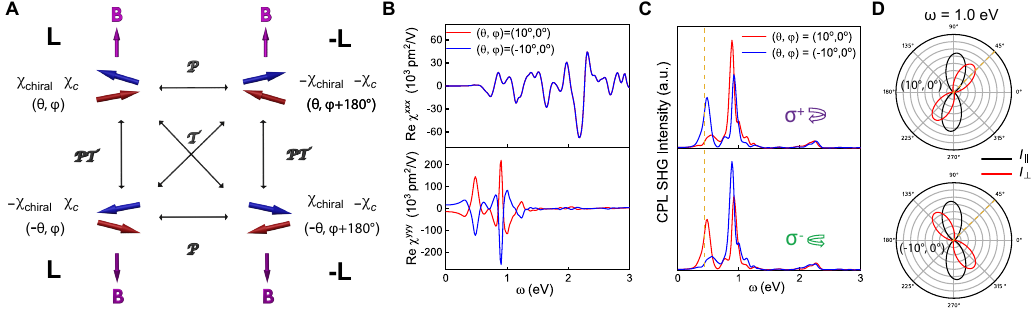}
    \caption{\textbf{Interference effects of chiral SHG in BL cAFM CrSBr.} (A) Representative canting states related by different symmetry operations and corresponding influences on SHG susceptibilities. (B) The SHG susceptibilities, (C) SHG intensity under circularly polarized light (CPL), and (D) polarization-resolved SHG under linearly polarized light at $\omega$=1.0\,eV of $\mathcal{PT}$-related canting states ($\pm10^\circ$, $0^\circ$). $I_{||}$ and $I_{\bot}$ represent the intensities of parallel-polarized and cross-polarized SHG under the illumination of normally incident linearly polarized light  \cite{zhu_2021_nanoletters_CrSBr} and the yellow dashed lines are eye-guided lines for notable change of SHG strength and polarization.} \label{Fig4}
\end{figure*}

To reveal the unique dependence on the canting angles of these SHG susceptibilities, we established a phenomenological theory to expand susceptibilities based on the corresponding order parameters. Considering a $\mathbf{B}$ field applied within the $xz$-plane (see Fig. \ref{Fig2}B), the N\'{e}el vector only has a $\hat{y}$-component $L_y \propto \cos\theta \cos\varphi$ (see Eq. (\ref{order-parameter})), and a net magnetization is induced along the $\hat{z}$-axis $M_z\propto \sin\theta$.   
Although magnetization along the $\hat{x}$-axis, $M_x$, can also be induced and breaks $\mathcal{PT}$ symmetry, it alone corresponds to (0$^\circ$,$\varphi$) canting states and results in vanishing in-plane chiral SHG components $\chi^{yyy}, \chi^{yxx}$ and $\chi^{xxy}$, as shown in Table \ref{tab:sym}. 
Consequently, it is $L_y$ and $M_z$ that truly matter in breaking $\mathcal{PT}$, $\mathcal{C}_{2z}\mathcal{T}$, $\mathcal{C}_{2x}$, and $\mathcal{M}_{y}$ symmetries, and activating in-plane chiral SHG components $\chi^{yyy}, \chi^{yxx}$ and $\chi^{xxy}$. In contrast, the in-plane $c$-type MSHG components $\chi^{xxx}, \chi^{xxy}$ and $\chi^{yxy}$ require only a nonzero value for $L_y$.  
Generally, the chiral SHG and $c$-type MSHG should be functions of their order parameters $L_y$ and $M_z$, and can be expanded in series form as
\begin{equation}
    \begin{aligned}
        \chi_{\rm chiral} &\propto \sum_{ij}D_{ij}L_y^{2i+1}M_z^{2j+1} \,,\\
        \chi_{c} &\propto \sum_iC_iL_y^{2i+1} \,.
    \end{aligned}
\end{equation}
Here $i$ and $j$ are nonnegative integers, and $D_{ij}$ ($C_{i}$) are coefficients of different terms. 
According to the symmetry requirements of $\chi_{\rm chiral}$ and $\chi_c$ shown in Fig. \ref{Fig4}A, only odd terms of $L_y$ and $M_z$ survive.
If we only keep the leading linear term, the SHG susceptibilities become
\begin{equation}
    \begin{aligned}
        \chi_{\rm chiral} &\propto D_{00}L_yM_z \\
        &\propto \cos\varphi \cos\theta \sin\theta \\
        &\propto \sin2\theta \cos\varphi  \\
        &\propto \kappa_x\,, \\
        \chi_{c} &\propto C_0L_y \\
        &\propto \cos\theta \cos\varphi \,.
    \end{aligned}
\end{equation}
These angle dependencies agree well with our calculation results.
More remarkably, without assuming $\boldsymbol{\kappa}$ as an order parameter
in advance, we retrieved the chiral SHG is proportional to the $\hat{x}$-component of $\boldsymbol{\kappa}$, i.e. $\kappa_x$ (see Eq. (\ref{order-parameter})). 
Moreover, in the absence of chiral SHG, this phenomenological theory is also applicable to the processes of in-plane spin canting and N\'{e}el vector rotations, demonstrating the robustness of our theory and enabling the detection of N\'{e}el vector through $c$-type MSHG in CrSBr (see figs. S11 and S12 \cite{SI}). \textcolor{black}{Furthermore, we have conducted further investigations on BL CrI$_3$ under an in-plane magnetic field along the $\hat{x}$-axis and BL CrCl$_3$ under an out-of-plane magnetic field along the $\hat{z}$-axis, which validate the generality and transferability of our theoretical predictions in CrSBr (see section S2 \cite{SI}).}

%******************************************************************%
%******************************************************************%

More interestingly, the spin-canting structure also induces an electric polarization proportional to the chiral SHG. Specifically, our calculations show that the induced polarization is oriented exclusively along the $\hat{y}$-direction, as required by the $\mathcal{C}_{2y}\mathcal{T}$ symmetry, and is denoted as $P_y$. 
Furthermore, the S.I. of the chiral SHG component $|\chi^{yyy}|$ is proportional to $P_y$, both under the variation of $\theta$ within the $(\theta,0^\circ)$ canting states (Fig. \ref{Fig3}C) and the variation of $\varphi$ within the $(10^\circ,\varphi)$ canting states (Fig. \ref{Fig3}F). This direct proportionality is also applicable for other chiral SHG components (fig. S10 \cite{SI}). 
Consequently, $P_y \propto \chi_{\rm chiral} \propto \sin 2\theta \cos \varphi \propto \kappa_x$ suggests a strong connection between spin-canting-induced electric polarization and chiral SHG, with $\boldsymbol{\kappa}$ acting as a bridge at the microscopic level. 
Indeed, it also supports the speculation that the MSHG observed in spin-canted CuCrP$_2$S$_6$ \cite{2024_AM_chiral-SHG-in-CuCrP2S6} and spin-spiral NiI$_2$ \cite{NiI2_2021_nl, NiI2_2022_Nature} originates from magnetization-induced polarization. 
This further indicates the existence of chiral SHG in these systems.
As spin-canting-induced electric polarization can be expressed as $\mathbf{P}=\hat{z} \times (\mathbf{S_A} \times \mathbf{S_B}) = \kappa_x \hat{y}$, its microscopic origin may be regarded as an extension of KNB theory \cite{KNB,gKNB} to van der Waals layers. The strong correlation between chiral SHG and $\boldsymbol{\kappa}$ enables chiral SHG to serve as a sensitive probe for the dynamics of spin-canting-induced electric polarization and to characterize the ME coupling in CrSBr, as depicted in Figs. \ref{Fig3} (A and D). 

%--------------------------------------------------
 \subsection*{Giant interference effects of chiral SHG in CrSBr}
%--------------------------------------------------
 
In addition to the tunability of magnitude, the controllability of the sign of $\boldsymbol{\kappa}$ enables BL CrSBr to exhibit the unique interference discussed previously (see Fig. \ref{Fig1}B and Eq. (2)). Generally, the opposite magnetic field $\pm\mathbf{B}$, combined with the opposite AFM domains $\pm\mathbf{L}$ in CrSBr, results in four distinct canting states, which are finely woven together by $\mathcal{P}$, $\mathcal{T}$, and $\mathcal{PT}$ operations, giving rise to rich symmetry-related physics in SHG, as depicted in Fig. \ref{Fig4}A. 
Specifically, unique interference effects occur when $\mathbf{B}$ is flipped in either the $\mathbf{L}$ or $\mathbf{-L}$ domains. 
As illustrated in the upper and lower panels of Fig. \ref{Fig4}A and numerically verified in the example of ($\pm10^\circ$, $0^\circ$) in Fig. \ref{Fig4}B, the chiral $\chi^{yyy}$, as an $i$-type tensor, reverses its sign while the $c$-type $\chi^{xxx}$ remains unchanged under $\mathbf{B}$ flipping. 
These two states are related by $\mathcal{PT}$ operation and $\boldsymbol{\kappa}$ reversal, and they show unique interference effects.
In the normal incident configuration, these interference effects manifest as the mirror operation along the $\hat{x}$-axis to the polarization-resolved SHG (Fig. \ref{Fig4}D) and the intensity change of the circularly polarized SHG under $\boldsymbol{\kappa}$ reversal, as well as the SHG circular dichroism effect (Fig. \ref{Fig4}C) (see details in section S1.4 \cite{SI}). 
Remarkably, these interference effects between chiral SHG and $c$-type MSHG can be very prominent. This is exemplified by the on/off switch of the outgoing circularly/linearly polarized SHG under the reversal of spin-canting direction (yellow dashed lines in Figs. \ref{Fig4} (C and D)). 
Such phenomena are difficult to achieve between crystal SHG and $c$-type MSHG \cite{npj, 2024-pnas-NLMO, Toyoda-2023-prm,Wu-2023-ACSnano,Tokura_FeGaO3_2004}.
From practical aspects, this enables 2D CrSBr to serve as an atomic-thin optical switch and optical filter, as depicted in fig. S7A \cite{SI}.

In addition, these interference effects are insensitive to the reversal of AFM domains. 
As illustrated in Fig. \ref{Fig4}A, under the same $\mathbf{B}$ field, the canting states in the $+\mathbf{L}$ domain and in the $-\mathbf{L}$ domain are related by the $\mathcal{P}$ operation. Therefore, 
all SHG components reverse signs in opposite domains. 
Because a global minus sign between domains cannot be manifested in the intensity of SHG, $\pm\mathbf{L}$ domains show identical SHG signals and interference effects.
This characteristic could be an important advantage for device applications, as the preparation of a single domain is rendered unnecessary (as depicted in figs. S7 (B and C) \cite{SI}).

Moreover, interference effects induced by chiral SHG and $c$-type MSHG possess numerous advantages compared to conventional interference effects that occur between crystal SHG and $c$-type MSHG \cite{npj, 2024-pnas-NLMO, Toyoda-2023-prm, Wu-2023-ACSnano, Tokura_FeGaO3_2004}.
On the one hand, the chiral SHG in CrSBr can be tuned to exactly match the $c$-type MSHG at an arbitrary $\omega$ via a $\mathbf{B}$ field, thus inducing maximal interference strength in these unconventional interference effects. 
On the other hand, the magnetic field required to reverse $\boldsymbol{\kappa}$ is much weaker than the field required to flip spins,  making these unconventional interference effects easier to achieve and more energy-efficient.
These features surpass the previous strategy with conventional interference effects for designing magneto-optical devices \cite{npj}.

We conducted further investigations of interference effects in BL CrI$_3$ under an in-plane magnetic field and BL CrCl$_3$ under an out-of-plane magnetic field to validate the generality and transferability of our findings in CrSBr.
As expected, notable interference effects are also observed in CrI$_3$ and CrCl$_3$ (figs. S17 and S18 \cite{SI}). However, because of the lower symmetry in these cases, the interference effects are more complex compared to CrSBr.
Moreover, in a very recent experimental work, a giant modulation of polarization-resolved SHG patterns under the reversal of spin-canting direction has been observed in CuCrP$_2$S$_6$ \cite{2024_AM_chiral-SHG-in-CuCrP2S6}, which is very similar to our results (Fig. \ref{Fig4}D). 
Therefore, this observation may serve as an experimental verification of our theory.

%=================================
\section*{Discussion} 
%=================================

%beyond \textcolor{black}{chiral} NLMO
The extraordinary properties of chiral SHG in 2D CrSBr enable numerous potential applications beyond the scope of this article.
In addition to interference effects within the ED-SHG, the ability to precisely tune the chiral SHG to a tiny value unleashes strong interference between the ED-SHG and the magnetic-dipole SHG (MD-SHG), which is typically orders of magnitude smaller than the crystal SHG. 
As shown in section S1.6 \cite{SI}, these interference effects arise from the sign reversal of MD-SHG under the reversal of the light propagation direction, enabling the SHG diode effects \cite{toyoda_2021_SA_nonreciprocal,Xu_2024_NC_SHG-diode}. 
Beyond interference effects, the strong coupling and the precise relationship between the chiral SHG and the canting angle offer unique advantages in the detection of magnetic structures. 
This not only enables chiral SHG to characterize static magnetic configurations but also allows it to serve as a probe of magnon dynamics in CrSBr, similar to other advanced experimental techniques such as transient optical reflectivity \cite{xu_2023_nature-nanotech_CrSBr_tunable_exciton-magnon_coupling}.

Besides, very recent studies have demonstrated that excitonic effects, which inevitably occur in 2D semiconductors, can enhance SHG \cite{Quek_2024_prl_exciton-enhanced-SHG, Louie_2024_nl_exciton-enhanced-NLO, Diana_2024_NC_exciton-enhanced-HHG} and other nonlinear optical responses such as shift current \cite{Louie_2021_PNAS_exciton-enhanced-shiftcurrent}. 
These exciting findings indicate that the large chiral SHG predicted in our work may be further enhanced by excitonic effects. 
Ensured by symmetry, the core findings of our work remain resilient in the face of excitonic effects. 
The potential enhancement of chiral SHG by excitonic effects is worth investigating in the future. 

%Generalize
Moreover, the concept of chiral SHG can be greatly extended. 
In 2D CrSBr, the chiral SHG and its vector spin chirality $\mathbf{\kappa}$ are imprinted by an external magnetic field $\mathbf{B}$, resulting in their high tunability. 
Differently, various 2D magnets with intrinsic spin chirality have been found in recent experiments, such as the helimagnet 2D NiI$_2$ \cite{NiI2_2022_Nature,NiI2_2021_nl} and skyrmions in 2D CrVI$_6$ \cite{CrVI6_nat-phys}, which can also contribute to chiral SHG.
Due to their complicated magnetic interactions, the magnitude and tunability of their chiral SHG merit further investigation.

%summary%
In summary, our study introduces the concept of chiral SHG, arising from the breaking of $\mathcal{P}$ and $\mathcal{PT}$ symmetries by the vector spin chirality $\mathbf{\kappa}$ induced by a magnetic field.
The highly tunable chiral SHG, ranging from zero to a value that is one order of magnitude larger than its intrinsic $c$-type MSHG, is demonstrated in 2D CrSBr through spin canting. 
Microscopically, we revealed that the driving mechanisms of chiral SHG are SOC and interlayer coupling and found that the large and anisotropic chiral SHG in CrSBr can be attributed to the anisotropic optical absorption and the band-nesting effect. 
Remarkably, both our numerical results and phenomenological theory demonstrate the chiral SHG is proportional to spin chirality and spin-canting-induced electric polarization, while the intrinsic $c$-type MSHG proportional to the N\'{e}el vector. These features enable accurate control of SHG by tuning the canting angle or, conversely, the detection of spin chirality, N\'{e}el vector, and ME effects.
\textcolor{black}{In addition, we verified the generality of our findings in BL cAFM CrI$_3$ and CrCl$_3$.}
% \MY{Outlook, Impact of our work to further work.}
Our findings pave the way for both fundamental research on chiral SHG and the development of multifunctional optical devices controlled by magnetic fields. 
The generality of these findings suggests a broad range of AFM magnets awaiting future exploration.

%=================================
\section*{Materials and Methods}\label{Materials and Methods}
%=================================
%--------------------------------------------------
 \subsection*{First-principles calculations}
%--------------------------------------------------
First-principles calculations were carried out using the Vienna $Ab \ initio$ Simulation Package \cite{VASP} with spin-orbit coupling included. The exchange-correlation functional with the Perdew-Burke-Ernzerhof form \cite{PBE} and the projector augmented-wave method \cite{PAW} were employed. A Hubbard $U$ value of 3\,eV was applied to the 3$d$ orbitals of the magnetic ion Cr. In the case of layered materials, we utilized the DFT-D3 van der Waals correction without damping \cite{DFT-D3}. The cut-off energy of the plane-wave basis was set to be 500\,eV. The force convergence criterion was set to be 1\,meV/\AA. The total energy converged within $10^{-6}$ eV. The k-point sampling used was $11\times9\times1$, and a vacuum thickness exceeding 15 \AA\, was maintained in the calculations for 2D materials.
%--------------------------------------------------
 \subsection*{SHG calculations}
%--------------------------------------------------
%optics
We utilized the Wannier90 code \cite{wannier90} to construct the tight-binding Hamiltonian and compute optical responses based on the converged wavefunctions \cite{Wang_2017,Chen_2022}. 
For each layer of CrSBr, we obtained 44 localized orbitals by projecting the Bloch wave functions onto atomic-like Wannier functions (including Cr-$d$ and S/Br-$p$ orbitals). The formulas used in the calculations of SHG are summarized in section S3.2 \cite{SI}. A broadening factor \textcolor{black}{$\eta$} of 0.05\,eV was applied to the Dirac delta function \textcolor{black}{in these calculations, which equates to introducing a phenomenological decoherence time $\tau = 13$ fs \cite{Chen_2022,sharma_2023_nl_graphene-decohenrence,sharma_2023_SA_WSe2-decohenrence,2021_nl_graphene-decohenrence}. A more detailed discussion about quantum decoherence is summarized in section S1.3}. 
We found $100\times 100\times1$ k-mesh is enough to obtain the converged SHG susceptibilities. We used a $300\times 300\times1$ k-mesh for the $\mathbf{k}$-resolved SHG calculation. 
It is worth noting that though the integral value of $c$-type MSHG is zero in $\chi^{yyy}$, there can be nonzero value at each single $\mathbf{k}$ point. Therefore, the formula for $i$-type SHG (see section S3.2 \cite{SI}) was used to calculate the $\mathbf{k}$-resolved SHG $\chi^{yyy}$, excluding any nonzero $c$-type contributions at each $\mathbf{k}$ point.
The S.I. of the SHG were calculated with the integration limits from 0 to 3\,eV. 
The influence of scissors correction \cite{Sipe_2005_prb_scissors} to the SHG susceptibility was discussed in section S1.2 \cite{SI}.

% \bibliography{ref.bib}

\begin{thebibliography}{60}%
\makeatletter
\providecommand \@ifxundefined [1]{%
 \@ifx{#1\undefined}
}%
\providecommand \@ifnum [1]{%
 \ifnum #1\expandafter \@firstoftwo
 \else \expandafter \@secondoftwo
 \fi
}%
\providecommand \@ifx [1]{%
 \ifx #1\expandafter \@firstoftwo
 \else \expandafter \@secondoftwo
 \fi
}%
\providecommand \natexlab [1]{#1}%
\providecommand \enquote  [1]{``#1''}%
\providecommand \bibnamefont  [1]{#1}%
\providecommand \bibfnamefont [1]{#1}%
\providecommand \citenamefont [1]{#1}%
\providecommand \href@noop [0]{\@secondoftwo}%
\providecommand \href [0]{\begingroup \@sanitize@url \@href}%
\providecommand \@href[1]{\@@startlink{#1}\@@href}%
\providecommand \@@href[1]{\endgroup#1\@@endlink}%
\providecommand \@sanitize@url [0]{\catcode `\\12\catcode `\$12\catcode `\&12\catcode `\#12\catcode `\^12\catcode `\_12\catcode `\%12\relax}%
\providecommand \@@startlink[1]{}%
\providecommand \@@endlink[0]{}%
\providecommand \url  [0]{\begingroup\@sanitize@url \@url }%
\providecommand \@url [1]{\endgroup\@href {#1}{\urlprefix }}%
\providecommand \urlprefix  [0]{URL }%
\providecommand \Eprint [0]{\href }%
\providecommand \doibase [0]{https://doi.org/}%
\providecommand \selectlanguage [0]{\@gobble}%
\providecommand \bibinfo  [0]{\@secondoftwo}%
\providecommand \bibfield  [0]{\@secondoftwo}%
\providecommand \translation [1]{[#1]}%
\providecommand \BibitemOpen [0]{}%
\providecommand \bibitemStop [0]{}%
\providecommand \bibitemNoStop [0]{.\EOS\space}%
\providecommand \EOS [0]{\spacefactor3000\relax}%
\providecommand \BibitemShut  [1]{\csname bibitem#1\endcsname}%
\let\auto@bib@innerbib\@empty
%</preamble>
\bibitem [{\citenamefont {Sun}\ \emph {et~al.}(2019)\citenamefont {Sun}, \citenamefont {Yi}, \citenamefont {Song}, \citenamefont {Clark}, \citenamefont {Huang}, \citenamefont {Shan}, \citenamefont {Wu}, \citenamefont {Huang}, \citenamefont {Gao}, \citenamefont {Chen}, \citenamefont {McGuire}, \citenamefont {Cao}, \citenamefont {Xiao}, \citenamefont {Liu}, \citenamefont {Yao}, \citenamefont {Xu},\ and\ \citenamefont {Wu}}]{Wu_CrI3_2019}%
  \BibitemOpen
  \bibfield  {author} {\bibinfo {author} {\bibfnamefont {Z.}~\bibnamefont {Sun}}, \bibinfo {author} {\bibfnamefont {Y.}~\bibnamefont {Yi}}, \bibinfo {author} {\bibfnamefont {T.}~\bibnamefont {Song}}, \bibinfo {author} {\bibfnamefont {G.}~\bibnamefont {Clark}}, \bibinfo {author} {\bibfnamefont {B.}~\bibnamefont {Huang}}, \bibinfo {author} {\bibfnamefont {Y.}~\bibnamefont {Shan}}, \bibinfo {author} {\bibfnamefont {S.}~\bibnamefont {Wu}}, \bibinfo {author} {\bibfnamefont {D.}~\bibnamefont {Huang}}, \bibinfo {author} {\bibfnamefont {C.}~\bibnamefont {Gao}}, \bibinfo {author} {\bibfnamefont {Z.}~\bibnamefont {Chen}}, \bibinfo {author} {\bibfnamefont {M.}~\bibnamefont {McGuire}}, \bibinfo {author} {\bibfnamefont {T.}~\bibnamefont {Cao}}, \bibinfo {author} {\bibfnamefont {D.}~\bibnamefont {Xiao}}, \bibinfo {author} {\bibfnamefont {W.-T.}\ \bibnamefont {Liu}}, \bibinfo {author} {\bibfnamefont {W.}~\bibnamefont {Yao}}, \bibinfo {author} {\bibfnamefont {X.}~\bibnamefont {Xu}},\ and\ \bibinfo {author} {\bibfnamefont
  {S.}~\bibnamefont {Wu}},\ }\bibfield  {title} {\bibinfo {title} {Giant nonreciprocal second-harmonic generation from antiferromagnetic bilayer {CrI$_3$}},\ }\href@noop {} {\bibfield  {journal} {\bibinfo  {journal} {Nature}\ }\textbf {\bibinfo {volume} {572}},\ \bibinfo {pages} {497} (\bibinfo {year} {2019})}\BibitemShut {NoStop}%
\bibitem [{\citenamefont {Chu}\ \emph {et~al.}(2020)\citenamefont {Chu}, \citenamefont {Roh}, \citenamefont {Island}, \citenamefont {Li}, \citenamefont {Lee}, \citenamefont {Chen}, \citenamefont {Park}, \citenamefont {Young}, \citenamefont {Lee},\ and\ \citenamefont {Hsieh}}]{MnPS3_2020_prl}%
  \BibitemOpen
  \bibfield  {author} {\bibinfo {author} {\bibfnamefont {H.}~\bibnamefont {Chu}}, \bibinfo {author} {\bibfnamefont {C.~J.}\ \bibnamefont {Roh}}, \bibinfo {author} {\bibfnamefont {J.~O.}\ \bibnamefont {Island}}, \bibinfo {author} {\bibfnamefont {C.}~\bibnamefont {Li}}, \bibinfo {author} {\bibfnamefont {S.}~\bibnamefont {Lee}}, \bibinfo {author} {\bibfnamefont {J.}~\bibnamefont {Chen}}, \bibinfo {author} {\bibfnamefont {J.-G.}\ \bibnamefont {Park}}, \bibinfo {author} {\bibfnamefont {A.~F.}\ \bibnamefont {Young}}, \bibinfo {author} {\bibfnamefont {J.~S.}\ \bibnamefont {Lee}},\ and\ \bibinfo {author} {\bibfnamefont {D.}~\bibnamefont {Hsieh}},\ }\bibfield  {title} {\bibinfo {title} {Linear magnetoelectric phase in ultrathin {MnPS$_3$} probed by optical second harmonic generation},\ }\href@noop {} {\bibfield  {journal} {\bibinfo  {journal} {Phys. Rev. Lett.}\ }\textbf {\bibinfo {volume} {124}},\ \bibinfo {pages} {027601} (\bibinfo {year} {2020})}\BibitemShut {NoStop}%
\bibitem [{\citenamefont {Lee}\ \emph {et~al.}(2021)\citenamefont {Lee}, \citenamefont {Dismukes}, \citenamefont {Telford}, \citenamefont {Wiscons}, \citenamefont {Wang}, \citenamefont {Xu}, \citenamefont {Nuckolls}, \citenamefont {Dean}, \citenamefont {Roy},\ and\ \citenamefont {Zhu}}]{zhu_2021_nanoletters_CrSBr}%
  \BibitemOpen
  \bibfield  {author} {\bibinfo {author} {\bibfnamefont {K.}~\bibnamefont {Lee}}, \bibinfo {author} {\bibfnamefont {A.~H.}\ \bibnamefont {Dismukes}}, \bibinfo {author} {\bibfnamefont {E.~J.}\ \bibnamefont {Telford}}, \bibinfo {author} {\bibfnamefont {R.~A.}\ \bibnamefont {Wiscons}}, \bibinfo {author} {\bibfnamefont {J.}~\bibnamefont {Wang}}, \bibinfo {author} {\bibfnamefont {X.}~\bibnamefont {Xu}}, \bibinfo {author} {\bibfnamefont {C.}~\bibnamefont {Nuckolls}}, \bibinfo {author} {\bibfnamefont {C.~R.}\ \bibnamefont {Dean}}, \bibinfo {author} {\bibfnamefont {X.}~\bibnamefont {Roy}},\ and\ \bibinfo {author} {\bibfnamefont {X.}~\bibnamefont {Zhu}},\ }\bibfield  {title} {\bibinfo {title} {Magnetic order and symmetry in the {2D} semiconductor {CrSBr}},\ }\href@noop {} {\bibfield  {journal} {\bibinfo  {journal} {Nano Lett.}\ }\textbf {\bibinfo {volume} {21}},\ \bibinfo {pages} {3511} (\bibinfo {year} {2021})}\BibitemShut {NoStop}%
\bibitem [{\citenamefont {Wu}\ \emph {et~al.}(2024)\citenamefont {Wu}, \citenamefont {Ye}, \citenamefont {Chen}, \citenamefont {Xu},\ and\ \citenamefont {Duan}}]{npj}%
  \BibitemOpen
  \bibfield  {author} {\bibinfo {author} {\bibfnamefont {D.}~\bibnamefont {Wu}}, \bibinfo {author} {\bibfnamefont {M.}~\bibnamefont {Ye}}, \bibinfo {author} {\bibfnamefont {H.}~\bibnamefont {Chen}}, \bibinfo {author} {\bibfnamefont {Y.}~\bibnamefont {Xu}},\ and\ \bibinfo {author} {\bibfnamefont {W.}~\bibnamefont {Duan}},\ }\bibfield  {title} {\bibinfo {title} {Giant and controllable nonlinear magneto-optical effects in two-dimensional magnets},\ }\href@noop {} {\bibfield  {journal} {\bibinfo  {journal} {npj Comput. Mater.}\ }\textbf {\bibinfo {volume} {10}},\ \bibinfo {pages} {79} (\bibinfo {year} {2024})}\BibitemShut {NoStop}%
\bibitem [{\citenamefont {Shoriki}\ \emph {et~al.}(2024)\citenamefont {Shoriki}, \citenamefont {Moriishi}, \citenamefont {Okamura}, \citenamefont {Yokoi}, \citenamefont {Usui}, \citenamefont {Murakawa}, \citenamefont {Sakai}, \citenamefont {Hanasaki}, \citenamefont {Tokura},\ and\ \citenamefont {Takahashi}}]{2024-pnas-NLMO}%
  \BibitemOpen
  \bibfield  {author} {\bibinfo {author} {\bibfnamefont {K.}~\bibnamefont {Shoriki}}, \bibinfo {author} {\bibfnamefont {K.}~\bibnamefont {Moriishi}}, \bibinfo {author} {\bibfnamefont {Y.}~\bibnamefont {Okamura}}, \bibinfo {author} {\bibfnamefont {K.}~\bibnamefont {Yokoi}}, \bibinfo {author} {\bibfnamefont {H.}~\bibnamefont {Usui}}, \bibinfo {author} {\bibfnamefont {H.}~\bibnamefont {Murakawa}}, \bibinfo {author} {\bibfnamefont {H.}~\bibnamefont {Sakai}}, \bibinfo {author} {\bibfnamefont {N.}~\bibnamefont {Hanasaki}}, \bibinfo {author} {\bibfnamefont {Y.}~\bibnamefont {Tokura}},\ and\ \bibinfo {author} {\bibfnamefont {Y.}~\bibnamefont {Takahashi}},\ }\bibfield  {title} {\bibinfo {title} {Large nonlinear optical magnetoelectric response in a noncentrosymmetric magnetic {Weyl} semimetal},\ }\href@noop {} {\bibfield  {journal} {\bibinfo  {journal} {Proc. Natl. Acad. Sci.}\ }\textbf {\bibinfo {volume} {121}},\ \bibinfo {pages} {e2316910121} (\bibinfo {year} {2024})}\BibitemShut {NoStop}%
\bibitem [{\citenamefont {Toyoda}\ \emph {et~al.}(2023)\citenamefont {Toyoda}, \citenamefont {Liao}, \citenamefont {Guo}, \citenamefont {Tokunaga}, \citenamefont {Arima}, \citenamefont {Tokura},\ and\ \citenamefont {Ogawa}}]{Toyoda-2023-prm}%
  \BibitemOpen
  \bibfield  {author} {\bibinfo {author} {\bibfnamefont {S.}~\bibnamefont {Toyoda}}, \bibinfo {author} {\bibfnamefont {J.-C.}\ \bibnamefont {Liao}}, \bibinfo {author} {\bibfnamefont {G.-Y.}\ \bibnamefont {Guo}}, \bibinfo {author} {\bibfnamefont {Y.}~\bibnamefont {Tokunaga}}, \bibinfo {author} {\bibfnamefont {T.-h.}\ \bibnamefont {Arima}}, \bibinfo {author} {\bibfnamefont {Y.}~\bibnamefont {Tokura}},\ and\ \bibinfo {author} {\bibfnamefont {N.}~\bibnamefont {Ogawa}},\ }\bibfield  {title} {\bibinfo {title} {Magnetic-field switching of second-harmonic generation in noncentrosymmetric magnet {${\mathrm{Eu}}_{2}\mathrm{Mn}{\mathrm{Si}}_{2}{\mathrm{O}}_{7}$}},\ }\href@noop {} {\bibfield  {journal} {\bibinfo  {journal} {Phys. Rev. Mater.}\ }\textbf {\bibinfo {volume} {7}},\ \bibinfo {pages} {024403} (\bibinfo {year} {2023})}\BibitemShut {NoStop}%
\bibitem [{\citenamefont {Wu}\ \emph {et~al.}(2023)\citenamefont {Wu}, \citenamefont {Fei}, \citenamefont {Sun}, \citenamefont {Yi}, \citenamefont {Xia}, \citenamefont {Yan}, \citenamefont {Guo}, \citenamefont {Shi}, \citenamefont {Yan}, \citenamefont {Cobden}, \citenamefont {Liu}, \citenamefont {Xu},\ and\ \citenamefont {Wu}}]{Wu-2023-ACSnano}%
  \BibitemOpen
  \bibfield  {author} {\bibinfo {author} {\bibfnamefont {S.}~\bibnamefont {Wu}}, \bibinfo {author} {\bibfnamefont {Z.}~\bibnamefont {Fei}}, \bibinfo {author} {\bibfnamefont {Z.}~\bibnamefont {Sun}}, \bibinfo {author} {\bibfnamefont {Y.}~\bibnamefont {Yi}}, \bibinfo {author} {\bibfnamefont {W.}~\bibnamefont {Xia}}, \bibinfo {author} {\bibfnamefont {D.}~\bibnamefont {Yan}}, \bibinfo {author} {\bibfnamefont {Y.}~\bibnamefont {Guo}}, \bibinfo {author} {\bibfnamefont {Y.}~\bibnamefont {Shi}}, \bibinfo {author} {\bibfnamefont {J.}~\bibnamefont {Yan}}, \bibinfo {author} {\bibfnamefont {D.~H.}\ \bibnamefont {Cobden}}, \bibinfo {author} {\bibfnamefont {W.-T.}\ \bibnamefont {Liu}}, \bibinfo {author} {\bibfnamefont {X.}~\bibnamefont {Xu}},\ and\ \bibinfo {author} {\bibfnamefont {S.}~\bibnamefont {Wu}},\ }\bibfield  {title} {\bibinfo {title} {Extrinsic nonlinear {Kerr} rotation in topological materials under a magnetic field},\ }\href@noop {} {\bibfield  {journal} {\bibinfo  {journal} {ACS nano}\ }\textbf {\bibinfo
  {volume} {17}},\ \bibinfo {pages} {18905} (\bibinfo {year} {2023})}\BibitemShut {NoStop}%
\bibitem [{\citenamefont {Ogawa}\ \emph {et~al.}(2004)\citenamefont {Ogawa}, \citenamefont {Kaneko}, \citenamefont {He}, \citenamefont {Yu}, \citenamefont {Arima},\ and\ \citenamefont {Tokura}}]{Tokura_FeGaO3_2004}%
  \BibitemOpen
  \bibfield  {author} {\bibinfo {author} {\bibfnamefont {Y.}~\bibnamefont {Ogawa}}, \bibinfo {author} {\bibfnamefont {Y.}~\bibnamefont {Kaneko}}, \bibinfo {author} {\bibfnamefont {J.~P.}\ \bibnamefont {He}}, \bibinfo {author} {\bibfnamefont {X.~Z.}\ \bibnamefont {Yu}}, \bibinfo {author} {\bibfnamefont {T.}~\bibnamefont {Arima}},\ and\ \bibinfo {author} {\bibfnamefont {Y.}~\bibnamefont {Tokura}},\ }\bibfield  {title} {\bibinfo {title} {Magnetization-induced second harmonic generation in a polar ferromagnet},\ }\href@noop {} {\bibfield  {journal} {\bibinfo  {journal} {Phys. Rev. Lett.}\ }\textbf {\bibinfo {volume} {92}},\ \bibinfo {pages} {047401} (\bibinfo {year} {2004})}\BibitemShut {NoStop}%
\bibitem [{\citenamefont {Lux}\ \emph {et~al.}(2020)\citenamefont {Lux}, \citenamefont {Freimuth}, \citenamefont {Bl{\"u}gel},\ and\ \citenamefont {Mokrousov}}]{chiral_Hall_2020_prl}%
  \BibitemOpen
  \bibfield  {author} {\bibinfo {author} {\bibfnamefont {F.~R.}\ \bibnamefont {Lux}}, \bibinfo {author} {\bibfnamefont {F.}~\bibnamefont {Freimuth}}, \bibinfo {author} {\bibfnamefont {S.}~\bibnamefont {Bl{\"u}gel}},\ and\ \bibinfo {author} {\bibfnamefont {Y.}~\bibnamefont {Mokrousov}},\ }\bibfield  {title} {\bibinfo {title} {Chiral {Hall} effect in noncollinear magnets from a cyclic cohomology approach},\ }\href@noop {} {\bibfield  {journal} {\bibinfo  {journal} {Phys. Rev. Lett.}\ }\textbf {\bibinfo {volume} {124}},\ \bibinfo {pages} {096602} (\bibinfo {year} {2020})}\BibitemShut {NoStop}%
\bibitem [{\citenamefont {Kipp}\ \emph {et~al.}(2021)\citenamefont {Kipp}, \citenamefont {Samanta}, \citenamefont {Lux}, \citenamefont {Merte}, \citenamefont {Go}, \citenamefont {Hanke}, \citenamefont {Redies}, \citenamefont {Freimuth}, \citenamefont {Blügel}, \citenamefont {Ležaić},\ and\ \citenamefont {Mokrousov}}]{chiral_Hall_2021_communications_physics}%
  \BibitemOpen
  \bibfield  {author} {\bibinfo {author} {\bibfnamefont {J.}~\bibnamefont {Kipp}}, \bibinfo {author} {\bibfnamefont {K.}~\bibnamefont {Samanta}}, \bibinfo {author} {\bibfnamefont {F.~R.}\ \bibnamefont {Lux}}, \bibinfo {author} {\bibfnamefont {M.}~\bibnamefont {Merte}}, \bibinfo {author} {\bibfnamefont {D.}~\bibnamefont {Go}}, \bibinfo {author} {\bibfnamefont {J.-P.}\ \bibnamefont {Hanke}}, \bibinfo {author} {\bibfnamefont {M.}~\bibnamefont {Redies}}, \bibinfo {author} {\bibfnamefont {F.}~\bibnamefont {Freimuth}}, \bibinfo {author} {\bibfnamefont {S.}~\bibnamefont {Blügel}}, \bibinfo {author} {\bibfnamefont {M.}~\bibnamefont {Ležaić}},\ and\ \bibinfo {author} {\bibfnamefont {Y.}~\bibnamefont {Mokrousov}},\ }\bibfield  {title} {\bibinfo {title} {The chiral {Hall} effect in canted ferromagnets and antiferromagnets},\ }\href@noop {} {\bibfield  {journal} {\bibinfo  {journal} {Commun. Phys.}\ }\textbf {\bibinfo {volume} {4}},\ \bibinfo {pages} {99} (\bibinfo {year} {2021})}\BibitemShut {NoStop}%
\bibitem [{\citenamefont {Zhou}\ \emph {et~al.}(2020)\citenamefont {Zhou}, \citenamefont {Hanke}, \citenamefont {Feng}, \citenamefont {Bl{\"u}gel}, \citenamefont {Mokrousov},\ and\ \citenamefont {Yao}}]{Yao_2020_prm_chiral-AHE&ANE}%
  \BibitemOpen
  \bibfield  {author} {\bibinfo {author} {\bibfnamefont {X.}~\bibnamefont {Zhou}}, \bibinfo {author} {\bibfnamefont {J.-P.}\ \bibnamefont {Hanke}}, \bibinfo {author} {\bibfnamefont {W.}~\bibnamefont {Feng}}, \bibinfo {author} {\bibfnamefont {S.}~\bibnamefont {Bl{\"u}gel}}, \bibinfo {author} {\bibfnamefont {Y.}~\bibnamefont {Mokrousov}},\ and\ \bibinfo {author} {\bibfnamefont {Y.}~\bibnamefont {Yao}},\ }\bibfield  {title} {\bibinfo {title} {Giant anomalous nernst effect in noncollinear antiferromagnetic {Mn}-based antiperovskite nitrides},\ }\href@noop {} {\bibfield  {journal} {\bibinfo  {journal} {Phys. Rev. Mater.}\ }\textbf {\bibinfo {volume} {4}},\ \bibinfo {pages} {024408} (\bibinfo {year} {2020})}\BibitemShut {NoStop}%
\bibitem [{\citenamefont {Bac}\ \emph {et~al.}(2022)\citenamefont {Bac}, \citenamefont {Koller}, \citenamefont {Lux}, \citenamefont {Wang}, \citenamefont {Riney}, \citenamefont {Borisiak}, \citenamefont {Powers}, \citenamefont {Zhukovskyi}, \citenamefont {Orlova}, \citenamefont {Dobrowolska}, \citenamefont {Furdyna}, \citenamefont {Dilley}, \citenamefont {Rokhinson}, \citenamefont {Mokrousov}, \citenamefont {McQueeney}, \citenamefont {Heinonen}, \citenamefont {Liu},\ and\ \citenamefont {Assaf}}]{chiral_Hall_2022_npjQM_MBT}%
  \BibitemOpen
  \bibfield  {author} {\bibinfo {author} {\bibfnamefont {S.-K.}\ \bibnamefont {Bac}}, \bibinfo {author} {\bibfnamefont {K.}~\bibnamefont {Koller}}, \bibinfo {author} {\bibfnamefont {F.}~\bibnamefont {Lux}}, \bibinfo {author} {\bibfnamefont {J.}~\bibnamefont {Wang}}, \bibinfo {author} {\bibfnamefont {L.}~\bibnamefont {Riney}}, \bibinfo {author} {\bibfnamefont {K.}~\bibnamefont {Borisiak}}, \bibinfo {author} {\bibfnamefont {W.}~\bibnamefont {Powers}}, \bibinfo {author} {\bibfnamefont {M.}~\bibnamefont {Zhukovskyi}}, \bibinfo {author} {\bibfnamefont {T.}~\bibnamefont {Orlova}}, \bibinfo {author} {\bibfnamefont {M.}~\bibnamefont {Dobrowolska}}, \bibinfo {author} {\bibfnamefont {J.~K.}\ \bibnamefont {Furdyna}}, \bibinfo {author} {\bibfnamefont {N.~R.}\ \bibnamefont {Dilley}}, \bibinfo {author} {\bibfnamefont {L.~P.}\ \bibnamefont {Rokhinson}}, \bibinfo {author} {\bibfnamefont {Y.}~\bibnamefont {Mokrousov}}, \bibinfo {author} {\bibfnamefont {R.~J.}\ \bibnamefont {McQueeney}}, \bibinfo {author} {\bibfnamefont
  {O.}~\bibnamefont {Heinonen}}, \bibinfo {author} {\bibfnamefont {X.}~\bibnamefont {Liu}},\ and\ \bibinfo {author} {\bibfnamefont {B.~A.}\ \bibnamefont {Assaf}},\ }\bibfield  {title} {\bibinfo {title} {Topological response of the anomalous {Hall} effect in {MnBi$_2$Te$_4$} due to magnetic canting},\ }\href@noop {} {\bibfield  {journal} {\bibinfo  {journal} {npj Quantum Mater.}\ }\textbf {\bibinfo {volume} {7}},\ \bibinfo {pages} {46} (\bibinfo {year} {2022})}\BibitemShut {NoStop}%
\bibitem [{\citenamefont {Cao}\ \emph {et~al.}(2023)\citenamefont {Cao}, \citenamefont {Jiang}, \citenamefont {Li}, \citenamefont {Tu}, \citenamefont {Zhou}, \citenamefont {Zhou},\ and\ \citenamefont {Yao}}]{Yao_2023_In-plane-anomalous-Hall}%
  \BibitemOpen
  \bibfield  {author} {\bibinfo {author} {\bibfnamefont {J.}~\bibnamefont {Cao}}, \bibinfo {author} {\bibfnamefont {W.}~\bibnamefont {Jiang}}, \bibinfo {author} {\bibfnamefont {X.-P.}\ \bibnamefont {Li}}, \bibinfo {author} {\bibfnamefont {D.}~\bibnamefont {Tu}}, \bibinfo {author} {\bibfnamefont {J.}~\bibnamefont {Zhou}}, \bibinfo {author} {\bibfnamefont {J.}~\bibnamefont {Zhou}},\ and\ \bibinfo {author} {\bibfnamefont {Y.}~\bibnamefont {Yao}},\ }\bibfield  {title} {\bibinfo {title} {In-plane anomalous {Hall} effect in {PT}-symmetric antiferromagnetic materials},\ }\href@noop {} {\bibfield  {journal} {\bibinfo  {journal} {Phys. Rev. Lett.}\ }\textbf {\bibinfo {volume} {130}},\ \bibinfo {pages} {166702} (\bibinfo {year} {2023})}\BibitemShut {NoStop}%
\bibitem [{\citenamefont {Zhou}\ \emph {et~al.}(2023)\citenamefont {Zhou}, \citenamefont {Feng}, \citenamefont {Li},\ and\ \citenamefont {Yao}}]{Yao_2023_nanolett_chiral-QAH&QTH}%
  \BibitemOpen
  \bibfield  {author} {\bibinfo {author} {\bibfnamefont {X.}~\bibnamefont {Zhou}}, \bibinfo {author} {\bibfnamefont {W.}~\bibnamefont {Feng}}, \bibinfo {author} {\bibfnamefont {Y.}~\bibnamefont {Li}},\ and\ \bibinfo {author} {\bibfnamefont {Y.}~\bibnamefont {Yao}},\ }\bibfield  {title} {\bibinfo {title} {Spin-chirality-driven quantum anomalous and quantum topological {Hall} effects in chiral magnets},\ }\href@noop {} {\bibfield  {journal} {\bibinfo  {journal} {Nano Lett.}\ } (\bibinfo {year} {2023})}\BibitemShut {NoStop}%
\bibitem [{\citenamefont {Merte}\ \emph {et~al.}(2023)\citenamefont {Merte}, \citenamefont {Freimuth}, \citenamefont {Go}, \citenamefont {Adamantopoulos}, \citenamefont {Lux}, \citenamefont {Plucinski}, \citenamefont {Gomonay}, \citenamefont {Bl{\"u}gel},\ and\ \citenamefont {Mokrousov}}]{chiral_photocurrents_2023_APL}%
  \BibitemOpen
  \bibfield  {author} {\bibinfo {author} {\bibfnamefont {M.}~\bibnamefont {Merte}}, \bibinfo {author} {\bibfnamefont {F.}~\bibnamefont {Freimuth}}, \bibinfo {author} {\bibfnamefont {D.}~\bibnamefont {Go}}, \bibinfo {author} {\bibfnamefont {T.}~\bibnamefont {Adamantopoulos}}, \bibinfo {author} {\bibfnamefont {F.}~\bibnamefont {Lux}}, \bibinfo {author} {\bibfnamefont {L.}~\bibnamefont {Plucinski}}, \bibinfo {author} {\bibfnamefont {O.}~\bibnamefont {Gomonay}}, \bibinfo {author} {\bibfnamefont {S.}~\bibnamefont {Bl{\"u}gel}},\ and\ \bibinfo {author} {\bibfnamefont {Y.}~\bibnamefont {Mokrousov}},\ }\bibfield  {title} {\bibinfo {title} {Photocurrents, inverse {Faraday} effect, and photospin {Hall} effect in {Mn$_2$Au}},\ }\href@noop {} {\bibfield  {journal} {\bibinfo  {journal} {APL Mater.}\ }\textbf {\bibinfo {volume} {11}} (\bibinfo {year} {2023})}\BibitemShut {NoStop}%
\bibitem [{\citenamefont {Ju}\ \emph {et~al.}(2021)\citenamefont {Ju}, \citenamefont {Lee}, \citenamefont {Kim}, \citenamefont {Choi}, \citenamefont {Roh}, \citenamefont {Son}, \citenamefont {Park}, \citenamefont {Kim}, \citenamefont {Jung}, \citenamefont {Kim}, \citenamefont {Kim}, \citenamefont {Park},\ and\ \citenamefont {Lee}}]{NiI2_2021_nl}%
  \BibitemOpen
  \bibfield  {author} {\bibinfo {author} {\bibfnamefont {H.}~\bibnamefont {Ju}}, \bibinfo {author} {\bibfnamefont {Y.}~\bibnamefont {Lee}}, \bibinfo {author} {\bibfnamefont {K.-T.}\ \bibnamefont {Kim}}, \bibinfo {author} {\bibfnamefont {I.~H.}\ \bibnamefont {Choi}}, \bibinfo {author} {\bibfnamefont {C.~J.}\ \bibnamefont {Roh}}, \bibinfo {author} {\bibfnamefont {S.}~\bibnamefont {Son}}, \bibinfo {author} {\bibfnamefont {P.}~\bibnamefont {Park}}, \bibinfo {author} {\bibfnamefont {J.~H.}\ \bibnamefont {Kim}}, \bibinfo {author} {\bibfnamefont {T.~S.}\ \bibnamefont {Jung}}, \bibinfo {author} {\bibfnamefont {J.~H.}\ \bibnamefont {Kim}}, \bibinfo {author} {\bibfnamefont {K.~H.}\ \bibnamefont {Kim}}, \bibinfo {author} {\bibfnamefont {J.-G.}\ \bibnamefont {Park}},\ and\ \bibinfo {author} {\bibfnamefont {J.~S.}\ \bibnamefont {Lee}},\ }\bibfield  {title} {\bibinfo {title} {Possible persistence of multiferroic order down to bilayer limit of van der waals material {NiI$_2$}},\ }\href@noop {} {\bibfield  {journal}
  {\bibinfo  {journal} {Nano lett.}\ }\textbf {\bibinfo {volume} {21}},\ \bibinfo {pages} {5126} (\bibinfo {year} {2021})}\BibitemShut {NoStop}%
\bibitem [{\citenamefont {Song}\ \emph {et~al.}(2022)\citenamefont {Song}, \citenamefont {Occhialini}, \citenamefont {Erge{\c{c}}en}, \citenamefont {Ilyas}, \citenamefont {Amoroso}, \citenamefont {Barone}, \citenamefont {Kapeghian}, \citenamefont {Watanabe}, \citenamefont {Taniguchi}, \citenamefont {Botana}, \citenamefont {Picozzi}, \citenamefont {Gedik},\ and\ \citenamefont {Comin}}]{NiI2_2022_Nature}%
  \BibitemOpen
  \bibfield  {author} {\bibinfo {author} {\bibfnamefont {Q.}~\bibnamefont {Song}}, \bibinfo {author} {\bibfnamefont {C.~A.}\ \bibnamefont {Occhialini}}, \bibinfo {author} {\bibfnamefont {E.}~\bibnamefont {Erge{\c{c}}en}}, \bibinfo {author} {\bibfnamefont {B.}~\bibnamefont {Ilyas}}, \bibinfo {author} {\bibfnamefont {D.}~\bibnamefont {Amoroso}}, \bibinfo {author} {\bibfnamefont {P.}~\bibnamefont {Barone}}, \bibinfo {author} {\bibfnamefont {J.}~\bibnamefont {Kapeghian}}, \bibinfo {author} {\bibfnamefont {K.}~\bibnamefont {Watanabe}}, \bibinfo {author} {\bibfnamefont {T.}~\bibnamefont {Taniguchi}}, \bibinfo {author} {\bibfnamefont {A.~S.}\ \bibnamefont {Botana}}, \bibinfo {author} {\bibfnamefont {S.}~\bibnamefont {Picozzi}}, \bibinfo {author} {\bibfnamefont {N.}~\bibnamefont {Gedik}},\ and\ \bibinfo {author} {\bibfnamefont {R.}~\bibnamefont {Comin}},\ }\bibfield  {title} {\bibinfo {title} {Evidence for a single-layer van der waals multiferroic},\ }\href@noop {} {\bibfield  {journal} {\bibinfo  {journal} {Nature}\
  }\textbf {\bibinfo {volume} {602}},\ \bibinfo {pages} {601} (\bibinfo {year} {2022})}\BibitemShut {NoStop}%
\bibitem [{\citenamefont {Aoki}\ \emph {et~al.}(2024)\citenamefont {Aoki}, \citenamefont {Dong}, \citenamefont {Wang}, \citenamefont {Huang}, \citenamefont {Itahashi}, \citenamefont {Ogawa}, \citenamefont {Ideue},\ and\ \citenamefont {Iwasa}}]{2024_AM_chiral-SHG-in-CuCrP2S6}%
  \BibitemOpen
  \bibfield  {author} {\bibinfo {author} {\bibfnamefont {S.}~\bibnamefont {Aoki}}, \bibinfo {author} {\bibfnamefont {Y.}~\bibnamefont {Dong}}, \bibinfo {author} {\bibfnamefont {Z.}~\bibnamefont {Wang}}, \bibinfo {author} {\bibfnamefont {X.~S.}\ \bibnamefont {Huang}}, \bibinfo {author} {\bibfnamefont {Y.~M.}\ \bibnamefont {Itahashi}}, \bibinfo {author} {\bibfnamefont {N.}~\bibnamefont {Ogawa}}, \bibinfo {author} {\bibfnamefont {T.}~\bibnamefont {Ideue}},\ and\ \bibinfo {author} {\bibfnamefont {Y.}~\bibnamefont {Iwasa}},\ }\bibfield  {title} {\bibinfo {title} {Giant modulation of the second harmonic generation by magnetoelectricity in two dimensional multiferroic {CuCrP$_2$S$_6$}.},\ }\href@noop {} {\bibfield  {journal} {\bibinfo  {journal} {Adv. Mater.}\ }\textbf {\bibinfo {volume} {36}},\ \bibinfo {pages} {2312781} (\bibinfo {year} {2024})}\BibitemShut {NoStop}%
\bibitem [{\citenamefont {Telford}\ \emph {et~al.}(2020)\citenamefont {Telford}, \citenamefont {Dismukes}, \citenamefont {Lee}, \citenamefont {Cheng}, \citenamefont {Wieteska}, \citenamefont {Bartholomew}, \citenamefont {Chen}, \citenamefont {Xu}, \citenamefont {Pasupathy}, \citenamefont {Zhu}, \citenamefont {Dean},\ and\ \citenamefont {Roy}}]{2020_AM_CrSBr}%
  \BibitemOpen
  \bibfield  {author} {\bibinfo {author} {\bibfnamefont {E.~J.}\ \bibnamefont {Telford}}, \bibinfo {author} {\bibfnamefont {A.~H.}\ \bibnamefont {Dismukes}}, \bibinfo {author} {\bibfnamefont {K.}~\bibnamefont {Lee}}, \bibinfo {author} {\bibfnamefont {M.}~\bibnamefont {Cheng}}, \bibinfo {author} {\bibfnamefont {A.}~\bibnamefont {Wieteska}}, \bibinfo {author} {\bibfnamefont {A.~K.}\ \bibnamefont {Bartholomew}}, \bibinfo {author} {\bibfnamefont {Y.-S.}\ \bibnamefont {Chen}}, \bibinfo {author} {\bibfnamefont {X.}~\bibnamefont {Xu}}, \bibinfo {author} {\bibfnamefont {A.~N.}\ \bibnamefont {Pasupathy}}, \bibinfo {author} {\bibfnamefont {X.}~\bibnamefont {Zhu}}, \bibinfo {author} {\bibfnamefont {C.~R.}\ \bibnamefont {Dean}},\ and\ \bibinfo {author} {\bibfnamefont {X.}~\bibnamefont {Roy}},\ }\bibfield  {title} {\bibinfo {title} {Layered antiferromagnetism induces large negative magnetoresistance in the van der {Waals} semiconductor {CrSBr}},\ }\href@noop {} {\bibfield  {journal} {\bibinfo  {journal} {Adv. Mater.}\
  }\textbf {\bibinfo {volume} {32}},\ \bibinfo {pages} {2003240} (\bibinfo {year} {2020})}\BibitemShut {NoStop}%
\bibitem [{\citenamefont {Wilson}\ \emph {et~al.}(2021)\citenamefont {Wilson}, \citenamefont {Lee}, \citenamefont {Cenker}, \citenamefont {Xie}, \citenamefont {Dismukes}, \citenamefont {Telford}, \citenamefont {Fonseca}, \citenamefont {Sivakumar}, \citenamefont {Dean}, \citenamefont {Cao}, \citenamefont {Roy}, \citenamefont {Xu},\ and\ \citenamefont {Zhu}}]{zhu_2021_nature-mater_CrSBr}%
  \BibitemOpen
  \bibfield  {author} {\bibinfo {author} {\bibfnamefont {N.~P.}\ \bibnamefont {Wilson}}, \bibinfo {author} {\bibfnamefont {K.}~\bibnamefont {Lee}}, \bibinfo {author} {\bibfnamefont {J.}~\bibnamefont {Cenker}}, \bibinfo {author} {\bibfnamefont {K.}~\bibnamefont {Xie}}, \bibinfo {author} {\bibfnamefont {A.~H.}\ \bibnamefont {Dismukes}}, \bibinfo {author} {\bibfnamefont {E.~J.}\ \bibnamefont {Telford}}, \bibinfo {author} {\bibfnamefont {J.}~\bibnamefont {Fonseca}}, \bibinfo {author} {\bibfnamefont {S.}~\bibnamefont {Sivakumar}}, \bibinfo {author} {\bibfnamefont {C.}~\bibnamefont {Dean}}, \bibinfo {author} {\bibfnamefont {T.}~\bibnamefont {Cao}}, \bibinfo {author} {\bibfnamefont {X.}~\bibnamefont {Roy}}, \bibinfo {author} {\bibfnamefont {X.}~\bibnamefont {Xu}},\ and\ \bibinfo {author} {\bibfnamefont {X.}~\bibnamefont {Zhu}},\ }\bibfield  {title} {\bibinfo {title} {Interlayer electronic coupling on demand in a {2D} magnetic semiconductor},\ }\href@noop {} {\bibfield  {journal} {\bibinfo  {journal} {Nat. Mater.}\
  }\textbf {\bibinfo {volume} {20}},\ \bibinfo {pages} {1657} (\bibinfo {year} {2021})}\BibitemShut {NoStop}%
\bibitem [{\citenamefont {Cenker}\ \emph {et~al.}(2022)\citenamefont {Cenker}, \citenamefont {Sivakumar}, \citenamefont {Xie}, \citenamefont {Miller}, \citenamefont {Thijssen}, \citenamefont {Liu}, \citenamefont {Dismukes}, \citenamefont {Fonseca}, \citenamefont {Anderson}, \citenamefont {Zhu}, \citenamefont {Roy}, \citenamefont {Xiao}, \citenamefont {Chu}, \citenamefont {Cao},\ and\ \citenamefont {Xu}}]{xu_2022_nature-nanotech_CrSBr}%
  \BibitemOpen
  \bibfield  {author} {\bibinfo {author} {\bibfnamefont {J.}~\bibnamefont {Cenker}}, \bibinfo {author} {\bibfnamefont {S.}~\bibnamefont {Sivakumar}}, \bibinfo {author} {\bibfnamefont {K.}~\bibnamefont {Xie}}, \bibinfo {author} {\bibfnamefont {A.}~\bibnamefont {Miller}}, \bibinfo {author} {\bibfnamefont {P.}~\bibnamefont {Thijssen}}, \bibinfo {author} {\bibfnamefont {Z.}~\bibnamefont {Liu}}, \bibinfo {author} {\bibfnamefont {A.~H.}\ \bibnamefont {Dismukes}}, \bibinfo {author} {\bibfnamefont {J.}~\bibnamefont {Fonseca}}, \bibinfo {author} {\bibfnamefont {E.}~\bibnamefont {Anderson}}, \bibinfo {author} {\bibfnamefont {X.}~\bibnamefont {Zhu}}, \bibinfo {author} {\bibfnamefont {X.}~\bibnamefont {Roy}}, \bibinfo {author} {\bibfnamefont {D.}~\bibnamefont {Xiao}}, \bibinfo {author} {\bibfnamefont {J.-H.}\ \bibnamefont {Chu}}, \bibinfo {author} {\bibfnamefont {T.}~\bibnamefont {Cao}},\ and\ \bibinfo {author} {\bibfnamefont {X.}~\bibnamefont {Xu}},\ }\bibfield  {title} {\bibinfo {title} {Reversible strain-induced
  magnetic phase transition in a van der {Waals} magnet},\ }\href@noop {} {\bibfield  {journal} {\bibinfo  {journal} {Nat. Nanotechnol.}\ }\textbf {\bibinfo {volume} {17}},\ \bibinfo {pages} {256} (\bibinfo {year} {2022})}\BibitemShut {NoStop}%
\bibitem [{\citenamefont {Bae}\ \emph {et~al.}(2022)\citenamefont {Bae}, \citenamefont {Wang}, \citenamefont {Scheie}, \citenamefont {Xu}, \citenamefont {Chica}, \citenamefont {Diederich}, \citenamefont {Cenker}, \citenamefont {Ziebel}, \citenamefont {Bai}, \citenamefont {Ren}, \citenamefont {Dean}, \citenamefont {Delor}, \citenamefont {Xu}, \citenamefont {Roy}, \citenamefont {Kent},\ and\ \citenamefont {Zhu}}]{zhu_2022_nature_CrSBr}%
  \BibitemOpen
  \bibfield  {author} {\bibinfo {author} {\bibfnamefont {Y.~J.}\ \bibnamefont {Bae}}, \bibinfo {author} {\bibfnamefont {J.}~\bibnamefont {Wang}}, \bibinfo {author} {\bibfnamefont {A.}~\bibnamefont {Scheie}}, \bibinfo {author} {\bibfnamefont {J.}~\bibnamefont {Xu}}, \bibinfo {author} {\bibfnamefont {D.~G.}\ \bibnamefont {Chica}}, \bibinfo {author} {\bibfnamefont {G.~M.}\ \bibnamefont {Diederich}}, \bibinfo {author} {\bibfnamefont {J.}~\bibnamefont {Cenker}}, \bibinfo {author} {\bibfnamefont {M.~E.}\ \bibnamefont {Ziebel}}, \bibinfo {author} {\bibfnamefont {Y.}~\bibnamefont {Bai}}, \bibinfo {author} {\bibfnamefont {H.}~\bibnamefont {Ren}}, \bibinfo {author} {\bibfnamefont {C.~R.}\ \bibnamefont {Dean}}, \bibinfo {author} {\bibfnamefont {M.}~\bibnamefont {Delor}}, \bibinfo {author} {\bibfnamefont {X.}~\bibnamefont {Xu}}, \bibinfo {author} {\bibfnamefont {X.}~\bibnamefont {Roy}}, \bibinfo {author} {\bibfnamefont {A.~D.}\ \bibnamefont {Kent}},\ and\ \bibinfo {author} {\bibfnamefont {X.}~\bibnamefont {Zhu}},\
  }\bibfield  {title} {\bibinfo {title} {Exciton-coupled coherent magnons in a {2D} semiconductor},\ }\href@noop {} {\bibfield  {journal} {\bibinfo  {journal} {Nature}\ }\textbf {\bibinfo {volume} {609}},\ \bibinfo {pages} {282} (\bibinfo {year} {2022})}\BibitemShut {NoStop}%
\bibitem [{\citenamefont {Ye}\ \emph {et~al.}(2022)\citenamefont {Ye}, \citenamefont {Wang}, \citenamefont {Wu}, \citenamefont {Liu}, \citenamefont {Zhou}, \citenamefont {Wang}, \citenamefont {Söll}, \citenamefont {Sofer}, \citenamefont {Yue}, \citenamefont {Liu}, \citenamefont {Tian}, \citenamefont {Xiong}, \citenamefont {Ji},\ and\ \citenamefont {Renshaw~Wang}}]{wang_2022_acsnano_CrSBr}%
  \BibitemOpen
  \bibfield  {author} {\bibinfo {author} {\bibfnamefont {C.}~\bibnamefont {Ye}}, \bibinfo {author} {\bibfnamefont {C.}~\bibnamefont {Wang}}, \bibinfo {author} {\bibfnamefont {Q.}~\bibnamefont {Wu}}, \bibinfo {author} {\bibfnamefont {S.}~\bibnamefont {Liu}}, \bibinfo {author} {\bibfnamefont {J.}~\bibnamefont {Zhou}}, \bibinfo {author} {\bibfnamefont {G.}~\bibnamefont {Wang}}, \bibinfo {author} {\bibfnamefont {A.}~\bibnamefont {Söll}}, \bibinfo {author} {\bibfnamefont {Z.}~\bibnamefont {Sofer}}, \bibinfo {author} {\bibfnamefont {M.}~\bibnamefont {Yue}}, \bibinfo {author} {\bibfnamefont {X.}~\bibnamefont {Liu}}, \bibinfo {author} {\bibfnamefont {M.}~\bibnamefont {Tian}}, \bibinfo {author} {\bibfnamefont {Q.}~\bibnamefont {Xiong}}, \bibinfo {author} {\bibfnamefont {W.}~\bibnamefont {Ji}},\ and\ \bibinfo {author} {\bibfnamefont {X.}~\bibnamefont {Renshaw~Wang}},\ }\bibfield  {title} {\bibinfo {title} {Layer-dependent interlayer antiferromagnetic spin reorientation in air-stable semiconductor {CrSBr}},\
  }\href@noop {} {\bibfield  {journal} {\bibinfo  {journal} {ACS nano}\ }\textbf {\bibinfo {volume} {16}},\ \bibinfo {pages} {11876} (\bibinfo {year} {2022})}\BibitemShut {NoStop}%
\bibitem [{\citenamefont {Diederich}\ \emph {et~al.}(2023)\citenamefont {Diederich}, \citenamefont {Cenker}, \citenamefont {Ren}, \citenamefont {Fonseca}, \citenamefont {Chica}, \citenamefont {Bae}, \citenamefont {Zhu}, \citenamefont {Roy}, \citenamefont {Cao}, \citenamefont {Xiao},\ and\ \citenamefont {Xu}}]{xu_2023_nature-nanotech_CrSBr_tunable_exciton-magnon_coupling}%
  \BibitemOpen
  \bibfield  {author} {\bibinfo {author} {\bibfnamefont {G.~M.}\ \bibnamefont {Diederich}}, \bibinfo {author} {\bibfnamefont {J.}~\bibnamefont {Cenker}}, \bibinfo {author} {\bibfnamefont {Y.}~\bibnamefont {Ren}}, \bibinfo {author} {\bibfnamefont {J.}~\bibnamefont {Fonseca}}, \bibinfo {author} {\bibfnamefont {D.~G.}\ \bibnamefont {Chica}}, \bibinfo {author} {\bibfnamefont {Y.~J.}\ \bibnamefont {Bae}}, \bibinfo {author} {\bibfnamefont {X.}~\bibnamefont {Zhu}}, \bibinfo {author} {\bibfnamefont {X.}~\bibnamefont {Roy}}, \bibinfo {author} {\bibfnamefont {T.}~\bibnamefont {Cao}}, \bibinfo {author} {\bibfnamefont {D.}~\bibnamefont {Xiao}},\ and\ \bibinfo {author} {\bibfnamefont {X.}~\bibnamefont {Xu}},\ }\bibfield  {title} {\bibinfo {title} {Tunable interaction between excitons and hybridized magnons in a layered semiconductor},\ }\href@noop {} {\bibfield  {journal} {\bibinfo  {journal} {Nat. Nanotechnol.}\ }\textbf {\bibinfo {volume} {18}},\ \bibinfo {pages} {23} (\bibinfo {year} {2023})}\BibitemShut {NoStop}%
\bibitem [{\citenamefont {Dirnberger}\ \emph {et~al.}(2023)\citenamefont {Dirnberger}, \citenamefont {Quan}, \citenamefont {Bushati}, \citenamefont {Diederich}, \citenamefont {Florian}, \citenamefont {Klein}, \citenamefont {Mosina}, \citenamefont {Sofer}, \citenamefont {Xu}, \citenamefont {Kamra}, \citenamefont {Garc{\'i}a-Vidal}, \citenamefont {Al{\`u}},\ and\ \citenamefont {Menon}}]{2023_Nature_CrSBr_tunable_MO}%
  \BibitemOpen
  \bibfield  {author} {\bibinfo {author} {\bibfnamefont {F.}~\bibnamefont {Dirnberger}}, \bibinfo {author} {\bibfnamefont {J.}~\bibnamefont {Quan}}, \bibinfo {author} {\bibfnamefont {R.}~\bibnamefont {Bushati}}, \bibinfo {author} {\bibfnamefont {G.~M.}\ \bibnamefont {Diederich}}, \bibinfo {author} {\bibfnamefont {M.}~\bibnamefont {Florian}}, \bibinfo {author} {\bibfnamefont {J.}~\bibnamefont {Klein}}, \bibinfo {author} {\bibfnamefont {K.}~\bibnamefont {Mosina}}, \bibinfo {author} {\bibfnamefont {Z.}~\bibnamefont {Sofer}}, \bibinfo {author} {\bibfnamefont {X.}~\bibnamefont {Xu}}, \bibinfo {author} {\bibfnamefont {A.}~\bibnamefont {Kamra}}, \bibinfo {author} {\bibfnamefont {F.~J.}\ \bibnamefont {Garc{\'i}a-Vidal}}, \bibinfo {author} {\bibfnamefont {A.}~\bibnamefont {Al{\`u}}},\ and\ \bibinfo {author} {\bibfnamefont {V.~M.}\ \bibnamefont {Menon}},\ }\bibfield  {title} {\bibinfo {title} {Magneto-optics in a van der {Waals} magnet tuned by self-hybridized polaritons},\ }\href@noop {} {\bibfield  {journal}
  {\bibinfo  {journal} {Nature}\ }\textbf {\bibinfo {volume} {620}},\ \bibinfo {pages} {533} (\bibinfo {year} {2023})}\BibitemShut {NoStop}%
\bibitem [{\citenamefont {Tabataba-Vakili}\ \emph {et~al.}(2024)\citenamefont {Tabataba-Vakili}, \citenamefont {Nguyen}, \citenamefont {Rupp}, \citenamefont {Mosina}, \citenamefont {Papavasileiou}, \citenamefont {Watanabe}, \citenamefont {Taniguchi}, \citenamefont {Maletinsky}, \citenamefont {Glazov}, \citenamefont {Sofer}, \citenamefont {Baimuratov},\ and\ \citenamefont {H{\"o}gele}}]{2024-NC-CrSBr-doping}%
  \BibitemOpen
  \bibfield  {author} {\bibinfo {author} {\bibfnamefont {F.}~\bibnamefont {Tabataba-Vakili}}, \bibinfo {author} {\bibfnamefont {H.~P.~G.}\ \bibnamefont {Nguyen}}, \bibinfo {author} {\bibfnamefont {A.}~\bibnamefont {Rupp}}, \bibinfo {author} {\bibfnamefont {K.}~\bibnamefont {Mosina}}, \bibinfo {author} {\bibfnamefont {A.}~\bibnamefont {Papavasileiou}}, \bibinfo {author} {\bibfnamefont {K.}~\bibnamefont {Watanabe}}, \bibinfo {author} {\bibfnamefont {T.}~\bibnamefont {Taniguchi}}, \bibinfo {author} {\bibfnamefont {P.}~\bibnamefont {Maletinsky}}, \bibinfo {author} {\bibfnamefont {M.~M.}\ \bibnamefont {Glazov}}, \bibinfo {author} {\bibfnamefont {Z.}~\bibnamefont {Sofer}}, \bibinfo {author} {\bibfnamefont {A.~S.}\ \bibnamefont {Baimuratov}},\ and\ \bibinfo {author} {\bibfnamefont {A.}~\bibnamefont {H{\"o}gele}},\ }\bibfield  {title} {\bibinfo {title} {Doping-control of excitons and magnetism in few-layer {CrSBr}},\ }\href@noop {} {\bibfield  {journal} {\bibinfo  {journal} {Nat. Commun.}\ }\textbf {\bibinfo {volume}
  {15}},\ \bibinfo {pages} {4735} (\bibinfo {year} {2024})}\BibitemShut {NoStop}%
\bibitem [{\citenamefont {Ziebel}\ \emph {et~al.}(2024)\citenamefont {Ziebel}, \citenamefont {Feuer}, \citenamefont {Cox}, \citenamefont {Zhu}, \citenamefont {Dean},\ and\ \citenamefont {Roy}}]{zhu_2024_CrSBr_nanoletter_review}%
  \BibitemOpen
  \bibfield  {author} {\bibinfo {author} {\bibfnamefont {M.~E.}\ \bibnamefont {Ziebel}}, \bibinfo {author} {\bibfnamefont {M.~L.}\ \bibnamefont {Feuer}}, \bibinfo {author} {\bibfnamefont {J.}~\bibnamefont {Cox}}, \bibinfo {author} {\bibfnamefont {X.}~\bibnamefont {Zhu}}, \bibinfo {author} {\bibfnamefont {C.~R.}\ \bibnamefont {Dean}},\ and\ \bibinfo {author} {\bibfnamefont {X.}~\bibnamefont {Roy}},\ }\bibfield  {title} {\bibinfo {title} {{CrSBr}: an air-stable, two-dimensional magnetic semiconductor},\ }\href@noop {} {\bibfield  {journal} {\bibinfo  {journal} {Nano Lett.}\ }\textbf {\bibinfo {volume} {24}},\ \bibinfo {pages} {4319} (\bibinfo {year} {2024})}\BibitemShut {NoStop}%
\bibitem [{\citenamefont {Wang}\ \emph {et~al.}(2017)\citenamefont {Wang}, \citenamefont {Liu}, \citenamefont {Kang}, \citenamefont {Gu}, \citenamefont {Xu},\ and\ \citenamefont {Duan}}]{Wang_2017}%
  \BibitemOpen
  \bibfield  {author} {\bibinfo {author} {\bibfnamefont {C.}~\bibnamefont {Wang}}, \bibinfo {author} {\bibfnamefont {X.}~\bibnamefont {Liu}}, \bibinfo {author} {\bibfnamefont {L.}~\bibnamefont {Kang}}, \bibinfo {author} {\bibfnamefont {B.-L.}\ \bibnamefont {Gu}}, \bibinfo {author} {\bibfnamefont {Y.}~\bibnamefont {Xu}},\ and\ \bibinfo {author} {\bibfnamefont {W.}~\bibnamefont {Duan}},\ }\bibfield  {title} {\bibinfo {title} {First-principles calculation of nonlinear optical responses by {Wannier} interpolation},\ }\href@noop {} {\bibfield  {journal} {\bibinfo  {journal} {Phys. Rev. B}\ }\textbf {\bibinfo {volume} {96}},\ \bibinfo {pages} {115147} (\bibinfo {year} {2017})}\BibitemShut {NoStop}%
\bibitem [{\citenamefont {Chen}\ \emph {et~al.}(2022)\citenamefont {Chen}, \citenamefont {Ye}, \citenamefont {Zou}, \citenamefont {Gu}, \citenamefont {Xu},\ and\ \citenamefont {Duan}}]{Chen_2022}%
  \BibitemOpen
  \bibfield  {author} {\bibinfo {author} {\bibfnamefont {H.}~\bibnamefont {Chen}}, \bibinfo {author} {\bibfnamefont {M.}~\bibnamefont {Ye}}, \bibinfo {author} {\bibfnamefont {N.}~\bibnamefont {Zou}}, \bibinfo {author} {\bibfnamefont {B.-L.}\ \bibnamefont {Gu}}, \bibinfo {author} {\bibfnamefont {Y.}~\bibnamefont {Xu}},\ and\ \bibinfo {author} {\bibfnamefont {W.}~\bibnamefont {Duan}},\ }\bibfield  {title} {\bibinfo {title} {Basic formulation and first-principles implementation of nonlinear magneto-optical effects},\ }\href@noop {} {\bibfield  {journal} {\bibinfo  {journal} {Phys. Rev. B}\ }\textbf {\bibinfo {volume} {105}},\ \bibinfo {pages} {075123} (\bibinfo {year} {2022})}\BibitemShut {NoStop}%
\bibitem [{\citenamefont {Guo}\ \emph {et~al.}(2023)\citenamefont {Guo}, \citenamefont {Qi}, \citenamefont {Zhang}, \citenamefont {Gao}, \citenamefont {Hu}, \citenamefont {Zhou}, \citenamefont {Zang}, \citenamefont {Zhao}, \citenamefont {Wang}, \citenamefont {Yan}, \citenamefont {Xu}, \citenamefont {Wu}, \citenamefont {Eda}, \citenamefont {Xiao}, \citenamefont {Yang}, \citenamefont {Gou}, \citenamefont {Feng}, \citenamefont {Guo}, \citenamefont {Zhou}, \citenamefont {Ren}, \citenamefont {Qiu}, \citenamefont {Pennycook},\ and\ \citenamefont {Wee}}]{NbOCl2_2023_nature}%
  \BibitemOpen
  \bibfield  {author} {\bibinfo {author} {\bibfnamefont {Q.}~\bibnamefont {Guo}}, \bibinfo {author} {\bibfnamefont {X.-Z.}\ \bibnamefont {Qi}}, \bibinfo {author} {\bibfnamefont {L.}~\bibnamefont {Zhang}}, \bibinfo {author} {\bibfnamefont {M.}~\bibnamefont {Gao}}, \bibinfo {author} {\bibfnamefont {S.}~\bibnamefont {Hu}}, \bibinfo {author} {\bibfnamefont {W.}~\bibnamefont {Zhou}}, \bibinfo {author} {\bibfnamefont {W.}~\bibnamefont {Zang}}, \bibinfo {author} {\bibfnamefont {X.}~\bibnamefont {Zhao}}, \bibinfo {author} {\bibfnamefont {J.}~\bibnamefont {Wang}}, \bibinfo {author} {\bibfnamefont {B.}~\bibnamefont {Yan}}, \bibinfo {author} {\bibfnamefont {M.}~\bibnamefont {Xu}}, \bibinfo {author} {\bibfnamefont {Y.-K.}\ \bibnamefont {Wu}}, \bibinfo {author} {\bibfnamefont {G.}~\bibnamefont {Eda}}, \bibinfo {author} {\bibfnamefont {Z.}~\bibnamefont {Xiao}}, \bibinfo {author} {\bibfnamefont {S.~A.}\ \bibnamefont {Yang}}, \bibinfo {author} {\bibfnamefont {H.}~\bibnamefont {Gou}}, \bibinfo {author} {\bibfnamefont {Y.~P.}\
  \bibnamefont {Feng}}, \bibinfo {author} {\bibfnamefont {G.-C.}\ \bibnamefont {Guo}}, \bibinfo {author} {\bibfnamefont {W.}~\bibnamefont {Zhou}}, \bibinfo {author} {\bibfnamefont {X.-F.}\ \bibnamefont {Ren}}, \bibinfo {author} {\bibfnamefont {C.-W.}\ \bibnamefont {Qiu}}, \bibinfo {author} {\bibfnamefont {S.~J.}\ \bibnamefont {Pennycook}},\ and\ \bibinfo {author} {\bibfnamefont {A.~T.~S.}\ \bibnamefont {Wee}},\ }\bibfield  {title} {\bibinfo {title} {Ultrathin quantum light source with van der {Waals} {NbOCl$_2$} crystal},\ }\href@noop {} {\bibfield  {journal} {\bibinfo  {journal} {Nature}\ }\textbf {\bibinfo {volume} {613}},\ \bibinfo {pages} {53} (\bibinfo {year} {2023})}\BibitemShut {NoStop}%
\bibitem [{\citenamefont {Yao}\ \emph {et~al.}(2021)\citenamefont {Yao}, \citenamefont {Finney}, \citenamefont {Zhang}, \citenamefont {Moore}, \citenamefont {Xian}, \citenamefont {Tancogne-Dejean}, \citenamefont {Liu}, \citenamefont {Ardelean}, \citenamefont {Xu}, \citenamefont {Halbertal}, \citenamefont {Watanabe}, \citenamefont {Taniguchi}, \citenamefont {Ochoa}, \citenamefont {Asenjo-Garcia}, \citenamefont {Zhu}, \citenamefont {Basov}, \citenamefont {Rubio}, \citenamefont {Dean}, \citenamefont {Hone},\ and\ \citenamefont {Schuck}}]{Yao_2021_SA_twisted-hBN}%
  \BibitemOpen
  \bibfield  {author} {\bibinfo {author} {\bibfnamefont {K.}~\bibnamefont {Yao}}, \bibinfo {author} {\bibfnamefont {N.~R.}\ \bibnamefont {Finney}}, \bibinfo {author} {\bibfnamefont {J.}~\bibnamefont {Zhang}}, \bibinfo {author} {\bibfnamefont {S.~L.}\ \bibnamefont {Moore}}, \bibinfo {author} {\bibfnamefont {L.}~\bibnamefont {Xian}}, \bibinfo {author} {\bibfnamefont {N.}~\bibnamefont {Tancogne-Dejean}}, \bibinfo {author} {\bibfnamefont {F.}~\bibnamefont {Liu}}, \bibinfo {author} {\bibfnamefont {J.}~\bibnamefont {Ardelean}}, \bibinfo {author} {\bibfnamefont {X.}~\bibnamefont {Xu}}, \bibinfo {author} {\bibfnamefont {D.}~\bibnamefont {Halbertal}}, \bibinfo {author} {\bibfnamefont {K.}~\bibnamefont {Watanabe}}, \bibinfo {author} {\bibfnamefont {T.}~\bibnamefont {Taniguchi}}, \bibinfo {author} {\bibfnamefont {H.}~\bibnamefont {Ochoa}}, \bibinfo {author} {\bibfnamefont {A.}~\bibnamefont {Asenjo-Garcia}}, \bibinfo {author} {\bibfnamefont {X.}~\bibnamefont {Zhu}}, \bibinfo {author} {\bibfnamefont {D.~N.}\ \bibnamefont
  {Basov}}, \bibinfo {author} {\bibfnamefont {A.}~\bibnamefont {Rubio}}, \bibinfo {author} {\bibfnamefont {C.~R.}\ \bibnamefont {Dean}}, \bibinfo {author} {\bibfnamefont {J.}~\bibnamefont {Hone}},\ and\ \bibinfo {author} {\bibfnamefont {P.~J.}\ \bibnamefont {Schuck}},\ }\bibfield  {title} {\bibinfo {title} {Enhanced tunable second harmonic generation from twistable interfaces and vertical superlattices in boron nitride homostructures},\ }\href@noop {} {\bibfield  {journal} {\bibinfo  {journal} {Sci. Adv.}\ }\textbf {\bibinfo {volume} {7}},\ \bibinfo {pages} {eabe8691} (\bibinfo {year} {2021})}\BibitemShut {NoStop}%
\bibitem [{\citenamefont {Wang}\ \emph {et~al.}(2024)\citenamefont {Wang}, \citenamefont {Ye}, \citenamefont {Guo}, \citenamefont {Li}, \citenamefont {Zou}, \citenamefont {Li}, \citenamefont {Zhang}, \citenamefont {Zhao}, \citenamefont {Xu}, \citenamefont {Chen}, \citenamefont {Wu}, \citenamefont {Bao}, \citenamefont {Xu},\ and\ \citenamefont {Duan}}]{wang_2024_prm}%
  \BibitemOpen
  \bibfield  {author} {\bibinfo {author} {\bibfnamefont {J.}~\bibnamefont {Wang}}, \bibinfo {author} {\bibfnamefont {M.}~\bibnamefont {Ye}}, \bibinfo {author} {\bibfnamefont {X.}~\bibnamefont {Guo}}, \bibinfo {author} {\bibfnamefont {Y.}~\bibnamefont {Li}}, \bibinfo {author} {\bibfnamefont {N.}~\bibnamefont {Zou}}, \bibinfo {author} {\bibfnamefont {H.}~\bibnamefont {Li}}, \bibinfo {author} {\bibfnamefont {Z.}~\bibnamefont {Zhang}}, \bibinfo {author} {\bibfnamefont {S.}~\bibnamefont {Zhao}}, \bibinfo {author} {\bibfnamefont {Z.}~\bibnamefont {Xu}}, \bibinfo {author} {\bibfnamefont {H.}~\bibnamefont {Chen}}, \bibinfo {author} {\bibfnamefont {D.}~\bibnamefont {Wu}}, \bibinfo {author} {\bibfnamefont {T.}~\bibnamefont {Bao}}, \bibinfo {author} {\bibfnamefont {Y.}~\bibnamefont {Xu}},\ and\ \bibinfo {author} {\bibfnamefont {W.}~\bibnamefont {Duan}},\ }\bibfield  {title} {\bibinfo {title} {Unbiased screening of deep-ultraviolet and mid-infrared nonlinear optical crystals: {Long-neglected} covalent and mixed-cation
  motifs},\ }\href@noop {} {\bibfield  {journal} {\bibinfo  {journal} {Phys. Rev. Mater.}\ }\textbf {\bibinfo {volume} {8}},\ \bibinfo {pages} {085202} (\bibinfo {year} {2024})}\BibitemShut {NoStop}%
\bibitem [{\citenamefont {Birss}(1964)}]{Birss_1964}%
  \BibitemOpen
  \bibfield  {author} {\bibinfo {author} {\bibfnamefont {R.~R.}\ \bibnamefont {Birss}},\ }\href@noop {} {\emph {\bibinfo {title} {Symmetry and magnetism}}},\ Vol.~\bibinfo {volume} {3}\ (\bibinfo  {publisher} {North-Holland Publishing Company},\ \bibinfo {year} {1964})\BibitemShut {NoStop}%
\bibitem [{\citenamefont {Shen}(1984)}]{shen1984principles}%
  \BibitemOpen
  \bibfield  {author} {\bibinfo {author} {\bibfnamefont {Y.-R.}\ \bibnamefont {Shen}},\ }\href@noop {} {\emph {\bibinfo {title} {Principles of nonlinear optics}}}\ (\bibinfo  {publisher} {Wiley-Interscience, New York, NY, USA},\ \bibinfo {year} {1984})\BibitemShut {NoStop}%
\bibitem [{SI()}]{SI}%
  \BibitemOpen
  \href@noop {} {\bibinfo {title} {Supporting information for ``spin-chirality driven second-harmonic generation in two-dimensional magnet {CrSBr}''}}\BibitemShut {NoStop}%
\bibitem [{\citenamefont {Xiao}\ \emph {et~al.}(2023)\citenamefont {Xiao}, \citenamefont {Shao}, \citenamefont {Gan}, \citenamefont {Wang}, \citenamefont {Han}, \citenamefont {Sheng}, \citenamefont {Zhang}, \citenamefont {Jiang},\ and\ \citenamefont {Li}}]{Xiao_classification}%
  \BibitemOpen
  \bibfield  {author} {\bibinfo {author} {\bibfnamefont {R.-C.}\ \bibnamefont {Xiao}}, \bibinfo {author} {\bibfnamefont {D.-F.}\ \bibnamefont {Shao}}, \bibinfo {author} {\bibfnamefont {W.}~\bibnamefont {Gan}}, \bibinfo {author} {\bibfnamefont {H.-W.}\ \bibnamefont {Wang}}, \bibinfo {author} {\bibfnamefont {H.}~\bibnamefont {Han}}, \bibinfo {author} {\bibfnamefont {Z.~G.}\ \bibnamefont {Sheng}}, \bibinfo {author} {\bibfnamefont {C.}~\bibnamefont {Zhang}}, \bibinfo {author} {\bibfnamefont {H.}~\bibnamefont {Jiang}},\ and\ \bibinfo {author} {\bibfnamefont {H.}~\bibnamefont {Li}},\ }\bibfield  {title} {\bibinfo {title} {Classification of second harmonic generation effect in magnetically ordered materials},\ }\href@noop {} {\bibfield  {journal} {\bibinfo  {journal} {npj Quantum Mater.}\ }\textbf {\bibinfo {volume} {8}},\ \bibinfo {pages} {62} (\bibinfo {year} {2023})}\BibitemShut {NoStop}%
\bibitem [{\citenamefont {Katsura}\ \emph {et~al.}(2005)\citenamefont {Katsura}, \citenamefont {Nagaosa},\ and\ \citenamefont {Balatsky}}]{KNB}%
  \BibitemOpen
  \bibfield  {author} {\bibinfo {author} {\bibfnamefont {H.}~\bibnamefont {Katsura}}, \bibinfo {author} {\bibfnamefont {N.}~\bibnamefont {Nagaosa}},\ and\ \bibinfo {author} {\bibfnamefont {A.~V.}\ \bibnamefont {Balatsky}},\ }\bibfield  {title} {\bibinfo {title} {Spin current and magnetoelectric effect in noncollinear magnets},\ }\href@noop {} {\bibfield  {journal} {\bibinfo  {journal} {Phys. Rev. Lett.}\ }\textbf {\bibinfo {volume} {95}},\ \bibinfo {pages} {057205} (\bibinfo {year} {2005})}\BibitemShut {NoStop}%
\bibitem [{\citenamefont {Xiang}\ \emph {et~al.}(2011)\citenamefont {Xiang}, \citenamefont {Kan}, \citenamefont {Zhang}, \citenamefont {Whangbo},\ and\ \citenamefont {Gong}}]{gKNB}%
  \BibitemOpen
  \bibfield  {author} {\bibinfo {author} {\bibfnamefont {H.}~\bibnamefont {Xiang}}, \bibinfo {author} {\bibfnamefont {E.}~\bibnamefont {Kan}}, \bibinfo {author} {\bibfnamefont {Y.}~\bibnamefont {Zhang}}, \bibinfo {author} {\bibfnamefont {M.-H.}\ \bibnamefont {Whangbo}},\ and\ \bibinfo {author} {\bibfnamefont {X.}~\bibnamefont {Gong}},\ }\bibfield  {title} {\bibinfo {title} {General theory for the ferroelectric polarization induced by spin-spiral order},\ }\href@noop {} {\bibfield  {journal} {\bibinfo  {journal} {Phys. Rev. Lett.}\ }\textbf {\bibinfo {volume} {107}},\ \bibinfo {pages} {157202} (\bibinfo {year} {2011})}\BibitemShut {NoStop}%
\bibitem [{\citenamefont {Ye}\ \emph {et~al.}(2023)\citenamefont {Ye}, \citenamefont {Zhou}, \citenamefont {Huang}, \citenamefont {Jiang}, \citenamefont {Guo}, \citenamefont {Cao}, \citenamefont {Yan}, \citenamefont {Wang}, \citenamefont {Jia}, \citenamefont {Jiang}, \citenamefont {Wang}, \citenamefont {Wu}, \citenamefont {Zhang}, \citenamefont {Li}, \citenamefont {Lei}, \citenamefont {Gou},\ and\ \citenamefont {Huang}}]{Huang_2023_NC_NbOCl}%
  \BibitemOpen
  \bibfield  {author} {\bibinfo {author} {\bibfnamefont {L.}~\bibnamefont {Ye}}, \bibinfo {author} {\bibfnamefont {W.}~\bibnamefont {Zhou}}, \bibinfo {author} {\bibfnamefont {D.}~\bibnamefont {Huang}}, \bibinfo {author} {\bibfnamefont {X.}~\bibnamefont {Jiang}}, \bibinfo {author} {\bibfnamefont {Q.}~\bibnamefont {Guo}}, \bibinfo {author} {\bibfnamefont {X.}~\bibnamefont {Cao}}, \bibinfo {author} {\bibfnamefont {S.}~\bibnamefont {Yan}}, \bibinfo {author} {\bibfnamefont {X.}~\bibnamefont {Wang}}, \bibinfo {author} {\bibfnamefont {D.}~\bibnamefont {Jia}}, \bibinfo {author} {\bibfnamefont {D.}~\bibnamefont {Jiang}}, \bibinfo {author} {\bibfnamefont {Y.}~\bibnamefont {Wang}}, \bibinfo {author} {\bibfnamefont {X.}~\bibnamefont {Wu}}, \bibinfo {author} {\bibfnamefont {X.}~\bibnamefont {Zhang}}, \bibinfo {author} {\bibfnamefont {Y.}~\bibnamefont {Li}}, \bibinfo {author} {\bibfnamefont {H.}~\bibnamefont {Lei}}, \bibinfo {author} {\bibfnamefont {H.}~\bibnamefont {Gou}},\ and\ \bibinfo {author} {\bibfnamefont
  {B.}~\bibnamefont {Huang}},\ }\bibfield  {title} {\bibinfo {title} {Manipulation of nonlinear optical responses in layered ferroelectric niobium oxide dihalides},\ }\href@noop {} {\bibfield  {journal} {\bibinfo  {journal} {Nat. Commun.}\ }\textbf {\bibinfo {volume} {14}},\ \bibinfo {pages} {5911} (\bibinfo {year} {2023})}\BibitemShut {NoStop}%
\bibitem [{\citenamefont {Carvalho}\ \emph {et~al.}(2013)\citenamefont {Carvalho}, \citenamefont {Ribeiro},\ and\ \citenamefont {Neto}}]{2013_prb_band-nesting}%
  \BibitemOpen
  \bibfield  {author} {\bibinfo {author} {\bibfnamefont {A.}~\bibnamefont {Carvalho}}, \bibinfo {author} {\bibfnamefont {R.}~\bibnamefont {Ribeiro}},\ and\ \bibinfo {author} {\bibfnamefont {A.~C.}\ \bibnamefont {Neto}},\ }\bibfield  {title} {\bibinfo {title} {Band nesting and the optical response of two-dimensional semiconducting transition metal dichalcogenides},\ }\href@noop {} {\bibfield  {journal} {\bibinfo  {journal} {Phys. Rev. B}\ }\textbf {\bibinfo {volume} {88}},\ \bibinfo {pages} {115205} (\bibinfo {year} {2013})}\BibitemShut {NoStop}%
\bibitem [{\citenamefont {Mennel}\ \emph {et~al.}(2020)\citenamefont {Mennel}, \citenamefont {Smejkal}, \citenamefont {Linhart}, \citenamefont {Burgd\"{o}rfer}, \citenamefont {Libisch},\ and\ \citenamefont {Mueller}}]{2020_nanolett_band-nesting}%
  \BibitemOpen
  \bibfield  {author} {\bibinfo {author} {\bibfnamefont {L.}~\bibnamefont {Mennel}}, \bibinfo {author} {\bibfnamefont {V.}~\bibnamefont {Smejkal}}, \bibinfo {author} {\bibfnamefont {L.}~\bibnamefont {Linhart}}, \bibinfo {author} {\bibfnamefont {J.}~\bibnamefont {Burgd\"{o}rfer}}, \bibinfo {author} {\bibfnamefont {F.}~\bibnamefont {Libisch}},\ and\ \bibinfo {author} {\bibfnamefont {T.}~\bibnamefont {Mueller}},\ }\bibfield  {title} {\bibinfo {title} {Band nesting in two-dimensional crystals: {An} exceptionally sensitive probe of strain},\ }\href@noop {} {\bibfield  {journal} {\bibinfo  {journal} {Nano Lett.}\ }\textbf {\bibinfo {volume} {20}},\ \bibinfo {pages} {4242} (\bibinfo {year} {2020})}\BibitemShut {NoStop}%
\bibitem [{\citenamefont {Klein}\ \emph {et~al.}(2023)\citenamefont {Klein}, \citenamefont {Pingault}, \citenamefont {Florian}, \citenamefont {Heißenbüttel}, \citenamefont {Steinhoff}, \citenamefont {Song}, \citenamefont {Torres}, \citenamefont {Dirnberger}, \citenamefont {Curtis}, \citenamefont {Weile}, \citenamefont {Penn}, \citenamefont {Deilmann}, \citenamefont {Dana}, \citenamefont {Bushati}, \citenamefont {Quan}, \citenamefont {Luxa}, \citenamefont {Sofer}, \citenamefont {Alù}, \citenamefont {Menon}, \citenamefont {Wurstbauer}, \citenamefont {Rohlfing}, \citenamefont {Narang}, \citenamefont {Lončar},\ and\ \citenamefont {Ross}}]{CrSBr_2023_1D_ACS-nano}%
  \BibitemOpen
  \bibfield  {author} {\bibinfo {author} {\bibfnamefont {J.}~\bibnamefont {Klein}}, \bibinfo {author} {\bibfnamefont {B.}~\bibnamefont {Pingault}}, \bibinfo {author} {\bibfnamefont {M.}~\bibnamefont {Florian}}, \bibinfo {author} {\bibfnamefont {M.-C.}\ \bibnamefont {Heißenbüttel}}, \bibinfo {author} {\bibfnamefont {A.}~\bibnamefont {Steinhoff}}, \bibinfo {author} {\bibfnamefont {Z.}~\bibnamefont {Song}}, \bibinfo {author} {\bibfnamefont {K.}~\bibnamefont {Torres}}, \bibinfo {author} {\bibfnamefont {F.}~\bibnamefont {Dirnberger}}, \bibinfo {author} {\bibfnamefont {J.~B.}\ \bibnamefont {Curtis}}, \bibinfo {author} {\bibfnamefont {M.}~\bibnamefont {Weile}}, \bibinfo {author} {\bibfnamefont {A.}~\bibnamefont {Penn}}, \bibinfo {author} {\bibfnamefont {T.}~\bibnamefont {Deilmann}}, \bibinfo {author} {\bibfnamefont {R.}~\bibnamefont {Dana}}, \bibinfo {author} {\bibfnamefont {R.}~\bibnamefont {Bushati}}, \bibinfo {author} {\bibfnamefont {J.}~\bibnamefont {Quan}}, \bibinfo {author} {\bibfnamefont {J.}~\bibnamefont
  {Luxa}}, \bibinfo {author} {\bibfnamefont {Z.}~\bibnamefont {Sofer}}, \bibinfo {author} {\bibfnamefont {A.}~\bibnamefont {Alù}}, \bibinfo {author} {\bibfnamefont {V.~M.}\ \bibnamefont {Menon}}, \bibinfo {author} {\bibfnamefont {U.}~\bibnamefont {Wurstbauer}}, \bibinfo {author} {\bibfnamefont {M.}~\bibnamefont {Rohlfing}}, \bibinfo {author} {\bibfnamefont {P.}~\bibnamefont {Narang}}, \bibinfo {author} {\bibfnamefont {M.}~\bibnamefont {Lončar}},\ and\ \bibinfo {author} {\bibfnamefont {F.~M.}\ \bibnamefont {Ross}},\ }\bibfield  {title} {\bibinfo {title} {The bulk van der {Waals} layered magnet {CrSBr} is a quasi-{1D} material},\ }\href@noop {} {\bibfield  {journal} {\bibinfo  {journal} {ACS nano}\ }\textbf {\bibinfo {volume} {17}},\ \bibinfo {pages} {5316} (\bibinfo {year} {2023})}\BibitemShut {NoStop}%
\bibitem [{\citenamefont {Boyd}(2020)}]{Boyd_2020_nonlinear-opticals}%
  \BibitemOpen
  \bibfield  {author} {\bibinfo {author} {\bibfnamefont {R.~W.}\ \bibnamefont {Boyd}},\ }\href@noop {} {\emph {\bibinfo {title} {Nonlinear Optics}}}\ (\bibinfo  {publisher} {Academic},\ \bibinfo {address} {San Diego},\ \bibinfo {year} {2020})\BibitemShut {NoStop}%
\bibitem [{\citenamefont {Sun}\ \emph {et~al.}(2025)\citenamefont {Sun}, \citenamefont {Hong}, \citenamefont {Chen}, \citenamefont {Sheng}, \citenamefont {Wu}, \citenamefont {Wang}, \citenamefont {Liang}, \citenamefont {Liu}, \citenamefont {Yuan}, \citenamefont {Wu}, \citenamefont {Mi}, \citenamefont {Liu}, \citenamefont {Shen},\ and\ \citenamefont {Wu}}]{Wu_2025_Nat-Mater_CrSBr-polymorphs}%
  \BibitemOpen
  \bibfield  {author} {\bibinfo {author} {\bibfnamefont {Z.}~\bibnamefont {Sun}}, \bibinfo {author} {\bibfnamefont {C.}~\bibnamefont {Hong}}, \bibinfo {author} {\bibfnamefont {Y.}~\bibnamefont {Chen}}, \bibinfo {author} {\bibfnamefont {Z.}~\bibnamefont {Sheng}}, \bibinfo {author} {\bibfnamefont {S.}~\bibnamefont {Wu}}, \bibinfo {author} {\bibfnamefont {Z.}~\bibnamefont {Wang}}, \bibinfo {author} {\bibfnamefont {B.}~\bibnamefont {Liang}}, \bibinfo {author} {\bibfnamefont {W.-T.}\ \bibnamefont {Liu}}, \bibinfo {author} {\bibfnamefont {Z.}~\bibnamefont {Yuan}}, \bibinfo {author} {\bibfnamefont {Y.}~\bibnamefont {Wu}}, \bibinfo {author} {\bibfnamefont {Q.}~\bibnamefont {Mi}}, \bibinfo {author} {\bibfnamefont {Z.}~\bibnamefont {Liu}}, \bibinfo {author} {\bibfnamefont {J.}~\bibnamefont {Shen}},\ and\ \bibinfo {author} {\bibfnamefont {S.}~\bibnamefont {Wu}},\ }\bibfield  {title} {\bibinfo {title} {Resolving and routing magnetic polymorphs in a {2D} layered antiferromagnet},\ }\href@noop {} {\bibfield  {journal}
  {\bibinfo  {journal} {Nat. Mater.}\ ,\ \bibinfo {pages} {1}} (\bibinfo {year} {2025})}\BibitemShut {NoStop}%
\bibitem [{\citenamefont {Toyoda}\ \emph {et~al.}(2021)\citenamefont {Toyoda}, \citenamefont {Fiebig}, \citenamefont {Arima}, \citenamefont {Tokura},\ and\ \citenamefont {Ogawa}}]{toyoda_2021_SA_nonreciprocal}%
  \BibitemOpen
  \bibfield  {author} {\bibinfo {author} {\bibfnamefont {S.}~\bibnamefont {Toyoda}}, \bibinfo {author} {\bibfnamefont {M.}~\bibnamefont {Fiebig}}, \bibinfo {author} {\bibfnamefont {T.-h.}\ \bibnamefont {Arima}}, \bibinfo {author} {\bibfnamefont {Y.}~\bibnamefont {Tokura}},\ and\ \bibinfo {author} {\bibfnamefont {N.}~\bibnamefont {Ogawa}},\ }\bibfield  {title} {\bibinfo {title} {Nonreciprocal second harmonic generation in a magnetoelectric material},\ }\href@noop {} {\bibfield  {journal} {\bibinfo  {journal} {Sci. Adv.}\ }\textbf {\bibinfo {volume} {7}},\ \bibinfo {pages} {eabe2793} (\bibinfo {year} {2021})}\BibitemShut {NoStop}%
\bibitem [{\citenamefont {Tzschaschel}\ \emph {et~al.}(2024)\citenamefont {Tzschaschel}, \citenamefont {Qiu}, \citenamefont {Gao}, \citenamefont {Li}, \citenamefont {Guo}, \citenamefont {Yang}, \citenamefont {Zhang}, \citenamefont {Xie}, \citenamefont {Liu}, \citenamefont {Gao}, \citenamefont {B{\'e}rub{\'e}}, \citenamefont {Dinh}, \citenamefont {Ho}, \citenamefont {Fang}, \citenamefont {Huang}, \citenamefont {Nordlander}, \citenamefont {Ma}, \citenamefont {Tafti}, \citenamefont {Moll}, \citenamefont {Law},\ and\ \citenamefont {Xu}}]{Xu_2024_NC_SHG-diode}%
  \BibitemOpen
  \bibfield  {author} {\bibinfo {author} {\bibfnamefont {C.}~\bibnamefont {Tzschaschel}}, \bibinfo {author} {\bibfnamefont {J.-X.}\ \bibnamefont {Qiu}}, \bibinfo {author} {\bibfnamefont {X.-J.}\ \bibnamefont {Gao}}, \bibinfo {author} {\bibfnamefont {H.-C.}\ \bibnamefont {Li}}, \bibinfo {author} {\bibfnamefont {C.}~\bibnamefont {Guo}}, \bibinfo {author} {\bibfnamefont {H.-Y.}\ \bibnamefont {Yang}}, \bibinfo {author} {\bibfnamefont {C.-P.}\ \bibnamefont {Zhang}}, \bibinfo {author} {\bibfnamefont {Y.-M.}\ \bibnamefont {Xie}}, \bibinfo {author} {\bibfnamefont {Y.-F.}\ \bibnamefont {Liu}}, \bibinfo {author} {\bibfnamefont {A.}~\bibnamefont {Gao}}, \bibinfo {author} {\bibfnamefont {D.}~\bibnamefont {B{\'e}rub{\'e}}}, \bibinfo {author} {\bibfnamefont {T.}~\bibnamefont {Dinh}}, \bibinfo {author} {\bibfnamefont {S.-C.}\ \bibnamefont {Ho}}, \bibinfo {author} {\bibfnamefont {Y.}~\bibnamefont {Fang}}, \bibinfo {author} {\bibfnamefont {F.}~\bibnamefont {Huang}}, \bibinfo {author} {\bibfnamefont {J.}~\bibnamefont
  {Nordlander}}, \bibinfo {author} {\bibfnamefont {Q.}~\bibnamefont {Ma}}, \bibinfo {author} {\bibfnamefont {F.}~\bibnamefont {Tafti}}, \bibinfo {author} {\bibfnamefont {P.~J.~W.}\ \bibnamefont {Moll}}, \bibinfo {author} {\bibfnamefont {K.~T.}\ \bibnamefont {Law}},\ and\ \bibinfo {author} {\bibfnamefont {S.-Y.}\ \bibnamefont {Xu}},\ }\bibfield  {title} {\bibinfo {title} {Nonlinear optical diode effect in a magnetic {Weyl} semimetal},\ }\href@noop {} {\bibfield  {journal} {\bibinfo  {journal} {Nat. Commun.}\ }\textbf {\bibinfo {volume} {15}},\ \bibinfo {pages} {3017} (\bibinfo {year} {2024})}\BibitemShut {NoStop}%
\bibitem [{\citenamefont {Xuan}\ \emph {et~al.}(2024)\citenamefont {Xuan}, \citenamefont {Lai}, \citenamefont {Wu},\ and\ \citenamefont {Quek}}]{Quek_2024_prl_exciton-enhanced-SHG}%
  \BibitemOpen
  \bibfield  {author} {\bibinfo {author} {\bibfnamefont {F.}~\bibnamefont {Xuan}}, \bibinfo {author} {\bibfnamefont {M.}~\bibnamefont {Lai}}, \bibinfo {author} {\bibfnamefont {Y.}~\bibnamefont {Wu}},\ and\ \bibinfo {author} {\bibfnamefont {S.~Y.}\ \bibnamefont {Quek}},\ }\bibfield  {title} {\bibinfo {title} {Exciton-enhanced spontaneous parametric down-conversion in two-dimensional crystals},\ }\href@noop {} {\bibfield  {journal} {\bibinfo  {journal} {Phys. Rev. Lett.}\ }\textbf {\bibinfo {volume} {132}},\ \bibinfo {pages} {246902} (\bibinfo {year} {2024})}\BibitemShut {NoStop}%
\bibitem [{\citenamefont {Ruan}\ \emph {et~al.}(2024)\citenamefont {Ruan}, \citenamefont {Chan},\ and\ \citenamefont {Louie}}]{Louie_2024_nl_exciton-enhanced-NLO}%
  \BibitemOpen
  \bibfield  {author} {\bibinfo {author} {\bibfnamefont {J.}~\bibnamefont {Ruan}}, \bibinfo {author} {\bibfnamefont {Y.-H.}\ \bibnamefont {Chan}},\ and\ \bibinfo {author} {\bibfnamefont {S.~G.}\ \bibnamefont {Louie}},\ }\bibfield  {title} {\bibinfo {title} {Exciton enhanced nonlinear optical responses in monolayer {h-BN} and {MoS$_2$}: Insight from first-principles exciton-state coupling formalism and calculations},\ }\href@noop {} {\bibfield  {journal} {\bibinfo  {journal} {Nano Lett.}\ }\textbf {\bibinfo {volume} {24}},\ \bibinfo {pages} {15533} (\bibinfo {year} {2024})}\BibitemShut {NoStop}%
\bibitem [{\citenamefont {Chang~Lee}\ \emph {et~al.}(2024)\citenamefont {Chang~Lee}, \citenamefont {Yue}, \citenamefont {Gaarde}, \citenamefont {Chan},\ and\ \citenamefont {Qiu}}]{Diana_2024_NC_exciton-enhanced-HHG}%
  \BibitemOpen
  \bibfield  {author} {\bibinfo {author} {\bibfnamefont {V.}~\bibnamefont {Chang~Lee}}, \bibinfo {author} {\bibfnamefont {L.}~\bibnamefont {Yue}}, \bibinfo {author} {\bibfnamefont {M.~B.}\ \bibnamefont {Gaarde}}, \bibinfo {author} {\bibfnamefont {Y.-h.}\ \bibnamefont {Chan}},\ and\ \bibinfo {author} {\bibfnamefont {D.~Y.}\ \bibnamefont {Qiu}},\ }\bibfield  {title} {\bibinfo {title} {Many-body enhancement of high-harmonic generation in monolayer {MoS$_2$}},\ }\href@noop {} {\bibfield  {journal} {\bibinfo  {journal} {Nat. Commun.}\ }\textbf {\bibinfo {volume} {15}},\ \bibinfo {pages} {6228} (\bibinfo {year} {2024})}\BibitemShut {NoStop}%
\bibitem [{\citenamefont {Chan}\ \emph {et~al.}(2021)\citenamefont {Chan}, \citenamefont {Qiu}, \citenamefont {da~Jornada},\ and\ \citenamefont {Louie}}]{Louie_2021_PNAS_exciton-enhanced-shiftcurrent}%
  \BibitemOpen
  \bibfield  {author} {\bibinfo {author} {\bibfnamefont {Y.-H.}\ \bibnamefont {Chan}}, \bibinfo {author} {\bibfnamefont {D.~Y.}\ \bibnamefont {Qiu}}, \bibinfo {author} {\bibfnamefont {F.~H.}\ \bibnamefont {da~Jornada}},\ and\ \bibinfo {author} {\bibfnamefont {S.~G.}\ \bibnamefont {Louie}},\ }\bibfield  {title} {\bibinfo {title} {Giant exciton-enhanced shift currents and direct current conduction with subbandgap photo excitations produced by many-electron interactions},\ }\href@noop {} {\bibfield  {journal} {\bibinfo  {journal} {Proc. Nat. Acad. Sci.}\ }\textbf {\bibinfo {volume} {118}},\ \bibinfo {pages} {e1906938118} (\bibinfo {year} {2021})}\BibitemShut {NoStop}%
\bibitem [{\citenamefont {Li}\ \emph {et~al.}(2024)\citenamefont {Li}, \citenamefont {Liu}, \citenamefont {Zhang}, \citenamefont {Zhang}, \citenamefont {Zhang}, \citenamefont {Zhang}, \citenamefont {Meng}, \citenamefont {Hou}, \citenamefont {Li}, \citenamefont {Kang}, \citenamefont {Huang}, \citenamefont {Cao}, \citenamefont {Hou}, \citenamefont {Cui}, \citenamefont {Zhang}, \citenamefont {Min}, \citenamefont {Lu}, \citenamefont {Xu}, \citenamefont {Sheng}, \citenamefont {Xiang},\ and\ \citenamefont {Zhang}}]{CrVI6_nat-phys}%
  \BibitemOpen
  \bibfield  {author} {\bibinfo {author} {\bibfnamefont {X.}~\bibnamefont {Li}}, \bibinfo {author} {\bibfnamefont {C.}~\bibnamefont {Liu}}, \bibinfo {author} {\bibfnamefont {Y.}~\bibnamefont {Zhang}}, \bibinfo {author} {\bibfnamefont {S.}~\bibnamefont {Zhang}}, \bibinfo {author} {\bibfnamefont {H.}~\bibnamefont {Zhang}}, \bibinfo {author} {\bibfnamefont {Y.}~\bibnamefont {Zhang}}, \bibinfo {author} {\bibfnamefont {W.}~\bibnamefont {Meng}}, \bibinfo {author} {\bibfnamefont {D.}~\bibnamefont {Hou}}, \bibinfo {author} {\bibfnamefont {T.}~\bibnamefont {Li}}, \bibinfo {author} {\bibfnamefont {C.}~\bibnamefont {Kang}}, \bibinfo {author} {\bibfnamefont {F.}~\bibnamefont {Huang}}, \bibinfo {author} {\bibfnamefont {R.}~\bibnamefont {Cao}}, \bibinfo {author} {\bibfnamefont {D.}~\bibnamefont {Hou}}, \bibinfo {author} {\bibfnamefont {P.}~\bibnamefont {Cui}}, \bibinfo {author} {\bibfnamefont {W.}~\bibnamefont {Zhang}}, \bibinfo {author} {\bibfnamefont {T.}~\bibnamefont {Min}}, \bibinfo {author} {\bibfnamefont
  {Q.}~\bibnamefont {Lu}}, \bibinfo {author} {\bibfnamefont {X.}~\bibnamefont {Xu}}, \bibinfo {author} {\bibfnamefont {Z.}~\bibnamefont {Sheng}}, \bibinfo {author} {\bibfnamefont {B.}~\bibnamefont {Xiang}},\ and\ \bibinfo {author} {\bibfnamefont {Z.}~\bibnamefont {Zhang}},\ }\bibfield  {title} {\bibinfo {title} {Topological {Kerr} effects in two-dimensional magnets with broken inversion symmetry},\ }\href@noop {} {\bibfield  {journal} {\bibinfo  {journal} {Nat. Phys.}\ }\textbf {\bibinfo {volume} {20}},\ \bibinfo {pages} {1145} (\bibinfo {year} {2024})}\BibitemShut {NoStop}%
\bibitem [{\citenamefont {Kresse}\ and\ \citenamefont {Furthm\"uller}(1996)}]{VASP}%
  \BibitemOpen
  \bibfield  {author} {\bibinfo {author} {\bibfnamefont {G.}~\bibnamefont {Kresse}}\ and\ \bibinfo {author} {\bibfnamefont {J.}~\bibnamefont {Furthm\"uller}},\ }\bibfield  {title} {\bibinfo {title} {Efficient iterative schemes for ab initio total-energy calculations using a plane-wave basis set},\ }\href@noop {} {\bibfield  {journal} {\bibinfo  {journal} {Phys. Rev. B}\ }\textbf {\bibinfo {volume} {54}},\ \bibinfo {pages} {11169} (\bibinfo {year} {1996})}\BibitemShut {NoStop}%
\bibitem [{\citenamefont {Perdew}\ \emph {et~al.}(1996)\citenamefont {Perdew}, \citenamefont {Burke},\ and\ \citenamefont {Ernzerhof}}]{PBE}%
  \BibitemOpen
  \bibfield  {author} {\bibinfo {author} {\bibfnamefont {J.~P.}\ \bibnamefont {Perdew}}, \bibinfo {author} {\bibfnamefont {K.}~\bibnamefont {Burke}},\ and\ \bibinfo {author} {\bibfnamefont {M.}~\bibnamefont {Ernzerhof}},\ }\bibfield  {title} {\bibinfo {title} {Generalized gradient approximation made simple},\ }\href@noop {} {\bibfield  {journal} {\bibinfo  {journal} {Phys. Rev. Lett.}\ }\textbf {\bibinfo {volume} {77}},\ \bibinfo {pages} {3865} (\bibinfo {year} {1996})}\BibitemShut {NoStop}%
\bibitem [{\citenamefont {Kresse}\ and\ \citenamefont {Joubert}(1999)}]{PAW}%
  \BibitemOpen
  \bibfield  {author} {\bibinfo {author} {\bibfnamefont {G.}~\bibnamefont {Kresse}}\ and\ \bibinfo {author} {\bibfnamefont {D.}~\bibnamefont {Joubert}},\ }\bibfield  {title} {\bibinfo {title} {From ultrasoft pseudopotentials to the projector augmented-wave method},\ }\href@noop {} {\bibfield  {journal} {\bibinfo  {journal} {Phys. Rev. B}\ }\textbf {\bibinfo {volume} {59}},\ \bibinfo {pages} {1758} (\bibinfo {year} {1999})}\BibitemShut {NoStop}%
\bibitem [{\citenamefont {Grimme}\ \emph {et~al.}(2010)\citenamefont {Grimme}, \citenamefont {Antony}, \citenamefont {Ehrlich},\ and\ \citenamefont {Krieg}}]{DFT-D3}%
  \BibitemOpen
  \bibfield  {author} {\bibinfo {author} {\bibfnamefont {S.}~\bibnamefont {Grimme}}, \bibinfo {author} {\bibfnamefont {J.}~\bibnamefont {Antony}}, \bibinfo {author} {\bibfnamefont {S.}~\bibnamefont {Ehrlich}},\ and\ \bibinfo {author} {\bibfnamefont {H.}~\bibnamefont {Krieg}},\ }\bibfield  {title} {\bibinfo {title} {A consistent and accurate ab initio parametrization of density functional dispersion correction {(DFT-D)} for the 94 elements {H-Pu}},\ }\href@noop {} {\bibfield  {journal} {\bibinfo  {journal} {J. Chem. Phys.}\ }\textbf {\bibinfo {volume} {132}},\ \bibinfo {pages} {154104} (\bibinfo {year} {2010})}\BibitemShut {NoStop}%
\bibitem [{\citenamefont {Mostofi}\ \emph {et~al.}(2014)\citenamefont {Mostofi}, \citenamefont {Yates}, \citenamefont {Pizzi}, \citenamefont {Lee}, \citenamefont {Souza}, \citenamefont {Vanderbilt},\ and\ \citenamefont {Marzari}}]{wannier90}%
  \BibitemOpen
  \bibfield  {author} {\bibinfo {author} {\bibfnamefont {A.~A.}\ \bibnamefont {Mostofi}}, \bibinfo {author} {\bibfnamefont {J.~R.}\ \bibnamefont {Yates}}, \bibinfo {author} {\bibfnamefont {G.}~\bibnamefont {Pizzi}}, \bibinfo {author} {\bibfnamefont {Y.-S.}\ \bibnamefont {Lee}}, \bibinfo {author} {\bibfnamefont {I.}~\bibnamefont {Souza}}, \bibinfo {author} {\bibfnamefont {D.}~\bibnamefont {Vanderbilt}},\ and\ \bibinfo {author} {\bibfnamefont {N.}~\bibnamefont {Marzari}},\ }\bibfield  {title} {\bibinfo {title} {An updated version of {Wannier90}: A tool for obtaining maximally-localised {Wannier} functions},\ }\href@noop {} {\bibfield  {journal} {\bibinfo  {journal} {Comput. Phys. Commun.}\ }\textbf {\bibinfo {volume} {185}},\ \bibinfo {pages} {2309} (\bibinfo {year} {2014})}\BibitemShut {NoStop}%
\bibitem [{\citenamefont {Sharma}\ \emph {et~al.}(2023{\natexlab{a}})\citenamefont {Sharma}, \citenamefont {Dewhurst},\ and\ \citenamefont {Shallcross}}]{sharma_2023_nl_graphene-decohenrence}%
  \BibitemOpen
  \bibfield  {author} {\bibinfo {author} {\bibfnamefont {S.}~\bibnamefont {Sharma}}, \bibinfo {author} {\bibfnamefont {J.~K.}\ \bibnamefont {Dewhurst}},\ and\ \bibinfo {author} {\bibfnamefont {S.}~\bibnamefont {Shallcross}},\ }\bibfield  {title} {\bibinfo {title} {Light-shaping of valley states},\ }\href@noop {} {\bibfield  {journal} {\bibinfo  {journal} {Nano Lett.}\ }\textbf {\bibinfo {volume} {23}},\ \bibinfo {pages} {11533} (\bibinfo {year} {2023}{\natexlab{a}})}\BibitemShut {NoStop}%
\bibitem [{\citenamefont {Sharma}\ \emph {et~al.}(2023{\natexlab{b}})\citenamefont {Sharma}, \citenamefont {Elliott},\ and\ \citenamefont {Shallcross}}]{sharma_2023_SA_WSe2-decohenrence}%
  \BibitemOpen
  \bibfield  {author} {\bibinfo {author} {\bibfnamefont {S.}~\bibnamefont {Sharma}}, \bibinfo {author} {\bibfnamefont {P.}~\bibnamefont {Elliott}},\ and\ \bibinfo {author} {\bibfnamefont {S.}~\bibnamefont {Shallcross}},\ }\bibfield  {title} {\bibinfo {title} {{THz} induced giant spin and valley currents},\ }\href@noop {} {\bibfield  {journal} {\bibinfo  {journal} {Sci. Adv.}\ }\textbf {\bibinfo {volume} {9}},\ \bibinfo {pages} {eadf3673} (\bibinfo {year} {2023}{\natexlab{b}})}\BibitemShut {NoStop}%
\bibitem [{\citenamefont {Heide}\ \emph {et~al.}(2021)\citenamefont {Heide}, \citenamefont {Eckstein}, \citenamefont {Boolakee}, \citenamefont {Gerner}, \citenamefont {Weber}, \citenamefont {Franco},\ and\ \citenamefont {Hommelhoff}}]{2021_nl_graphene-decohenrence}%
  \BibitemOpen
  \bibfield  {author} {\bibinfo {author} {\bibfnamefont {C.}~\bibnamefont {Heide}}, \bibinfo {author} {\bibfnamefont {T.}~\bibnamefont {Eckstein}}, \bibinfo {author} {\bibfnamefont {T.}~\bibnamefont {Boolakee}}, \bibinfo {author} {\bibfnamefont {C.}~\bibnamefont {Gerner}}, \bibinfo {author} {\bibfnamefont {H.~B.}\ \bibnamefont {Weber}}, \bibinfo {author} {\bibfnamefont {I.}~\bibnamefont {Franco}},\ and\ \bibinfo {author} {\bibfnamefont {P.}~\bibnamefont {Hommelhoff}},\ }\bibfield  {title} {\bibinfo {title} {Electronic coherence and coherent dephasing in the optical control of electrons in graphene},\ }\href@noop {} {\bibfield  {journal} {\bibinfo  {journal} {Nano Lett.}\ }\textbf {\bibinfo {volume} {21}},\ \bibinfo {pages} {9403} (\bibinfo {year} {2021})}\BibitemShut {NoStop}%
\bibitem [{\citenamefont {Nastos}\ \emph {et~al.}(2005)\citenamefont {Nastos}, \citenamefont {Olejnik}, \citenamefont {Schwarz},\ and\ \citenamefont {Sipe}}]{Sipe_2005_prb_scissors}%
  \BibitemOpen
  \bibfield  {author} {\bibinfo {author} {\bibfnamefont {F.}~\bibnamefont {Nastos}}, \bibinfo {author} {\bibfnamefont {B.}~\bibnamefont {Olejnik}}, \bibinfo {author} {\bibfnamefont {K.}~\bibnamefont {Schwarz}},\ and\ \bibinfo {author} {\bibfnamefont {J.}~\bibnamefont {Sipe}},\ }\bibfield  {title} {\bibinfo {title} {Scissors implementation within length-gauge formulations of the frequency-dependent nonlinear optical response of semiconductors},\ }\href@noop {} {\bibfield  {journal} {\bibinfo  {journal} {Phys. Rev. B}\ }\textbf {\bibinfo {volume} {72}},\ \bibinfo {pages} {045223} (\bibinfo {year} {2005})}\BibitemShut {NoStop}%
\end{thebibliography}
%

\section*{Acknowledgments}
The work was carried out at the National Supercomputer Center in Tianjin using the Tianhe new-generation supercomputer.\\
\textbf{Funding:}
This work was supported by the Basic Science Center Project of NSFC (grant no. 52388201), the Ministry of Science and Technology of China (grant no. 2023YFA1406400), the National Natural Science Foundation of China (grants no. 12334003, no. 12421004, and no. 12361141826), the Innovation Program for Quantum Science and Technology (Grant No. 2023ZD0300500), NSAF (Grant No. U2330401), and the National Science Fund for Distinguished Young Scholars (grant no. 12025405). \\ 
\textbf{Author contributions:}
D.W., M.Y. and Y.X. conceived the project. D.W. carried out the numerical calculations and analyzed the data. D.W. and M.Y. discussed the results and wrote the manuscript with input from all other authors. M.Y., Y.X. and W.D. supervised the project.\\
\textbf{Competing interests:}
The authors declare that they have no competing interests. \\
\textbf{Data and materials availability:}
All data needed to evaluate the conclusions in the paper are present in the paper and/or the Supplementary Materials.
\end{document}